\documentclass[times,twocolumn]{aastex62}

\usepackage{CJK}
\usepackage{amsmath}
\usepackage{multirow}
\usepackage{cases}
\usepackage{graphicx}
\usepackage{subfigure}
\usepackage{natbib}
\usepackage{color}
\usepackage{tabularx}
\usepackage{bm}
\usepackage{threeparttable}
\usepackage{gensymb}

\newcommand{\thetae}{\theta_{\rm E}}

\newcommand{\pie}{\pi_{\rm E}}

\newcommand{\te}{t_{\rm E}}

\newcommand{\event}{KMT-2022-BLG-1818}

\newcommand{\hjd}{{\rm HJD}^{\prime}}
\DeclareGraphicsExtensions{.pdf,.png,.jpg}

\shorttitle{}
\shortauthors{Li et al.}

\begin{document}
\begin{CJK*}{UTF8}{gbsn}
\title{{\large KMT-2022-BLG-1818Lb,c: A Cold Super-Jupiter with a Saturn Sibling}}

\def\thefootnote{*}\footnotetext{H.L. and J.Z. contributed equally to this work.}\def\thefootnote{\arabic{footnote}}

\correspondingauthor{Weicheng Zang}
\email{zangweicheng@westlake.edu.cn}

\author{Hongyu Li$^*$}
\affiliation{Department of Astronomy, Tsinghua University, Beijing 100084, China}
\affiliation{Department of Astronomy, School of Physics, Huazhong University of Science and Technology, Wuhan, 430074, China}

\author[0000-0002-1279-0666]{Jiyuan Zhang$^*$}
\affiliation{Department of Astronomy, Tsinghua University, Beijing 100084, China}
%zhangjy22@mails.tsinghua.edu.cn

\author{Cheongho Han}
\affiliation{Department of Physics, Chungbuk National University, Cheongju 28644, Republic of Korea}
%cheongho@astroph.chungbuk.ac.kr

\author[0000-0001-6000-3463]{Weicheng Zang}
\affiliation{Center for Astrophysics $|$ Harvard \& Smithsonian, 60 Garden St., Cambridge, MA 02138, USA}

\author[0000-0002-0314-6000]{Youn Kil Jung}
\affiliation{Korea Astronomy and Space Science Institute, Daejeon 34055, Republic of Korea}
\affiliation{National University of Science and Technology (UST), Daejeon 34113, Republic of Korea}
%\email{ykjung21@kasi.re.kr}

\author[0000-0001-5207-5619]{Andrzej Udalski}
\affiliation{Astronomical Observatory, University of Warsaw, Al. Ujazdowskie 4, 00-478 Warszawa, Poland}
%udalski@astrouw.edu.pl

\author{Takahiro Sumi}
\affiliation{Department of Earth and Space Science, Graduate School of Science, Osaka University, Toyonaka, Osaka 560-0043, Japan}

\author[0000-0003-0626-8465]{Hongjing Yang}
\affiliation{Department of Astronomy, Westlake University, Hangzhou 310030, Zhejiang Province, China}
%yanghongjing@westlake.edu.cn

\author[0000-0003-2337-0533]{Renkun Kuang}
\affiliation{independent researcher}
%renkunkuang@gmail.com

\author[0000-0001-8317-2788]{Shude Mao}
\affiliation{Department of Astronomy, Westlake University, Hangzhou 310030, Zhejiang Province, China}
%shude.mao@gmail.com

\collaboration{(Leading Authors)}

\author[0000-0003-3316-4012]{Michael D. Albrow}
\affiliation{University of Canterbury, School of Physical and Chemical Sciences, Private Bag 4800, Christchurch 8020, New Zealand}
%\email{michael.albrow@canterbury.ac.nz}

\author[0000-0001-6285-4528]{Sun-Ju Chung}
\affiliation{Korea Astronomy and Space Science Institute, Daejeon 34055, Republic of Korea}
%\email{sjchung@kasi.re.kr}

\author{Andrew Gould} % No OrcID on purpose
\affiliation{Max-Planck-Institute for Astronomy, K\"onigstuhl 17, 69117 Heidelberg, Germany}
\affiliation{Department of Astronomy, Ohio State University, 140 W. 18th Ave., Columbus, OH 43210, USA}
%\email{gould.34@osu.edu}

\author[0000-0002-9241-4117]{Kyu-Ha Hwang}
\affiliation{Korea Astronomy and Space Science Institute, Daejeon 34055, Republic of Korea}
%\email{kyuha@kasi.re.kr}

\author[0000-0001-9823-2907]{Yoon-Hyun Ryu} 
\affiliation{Korea Astronomy and Space Science Institute, Daejeon 34055, Republic of Korea}
%yoonhyunryu@gmail.com

\author[0000-0002-4355-9838]{In-Gu Shin}
\affiliation{Center for Astrophysics $|$ Harvard \& Smithsonian, 60 Garden St.,Cambridge, MA 02138, USA}
%\email{ingushin@gmail.com}

\author[0000-0003-1525-5041]{Yossi Shvartzvald}
\affiliation{Department of Particle Physics and Astrophysics, Weizmann Institute of Science, Rehovot 7610001, Israel}
%\email{yossishv@gmail.com}

\author[0000-0001-9481-7123]{Jennifer C. Yee}
\affiliation{Center for Astrophysics $|$ Harvard \& Smithsonian, 60 Garden St.,Cambridge, MA 02138, USA}
%\email{jyee@cfa.harvard.edu}

\author[0000-0002-7511-2950]{Sang-Mok Cha} % ONLY for events with years <= 2023
\affiliation{Korea Astronomy and Space Science Institute, Daejeon 34055, Republic of Korea}
\affiliation{School of Space Research, Kyung Hee University, Yongin, Kyeonggi 17104, Republic of Korea} 
%\email{chasm@kasi.re.kr}

\author{Dong-Jin Kim}
\affiliation{Korea Astronomy and Space Science Institute, Daejeon 34055, Republic of Korea}
%\email{keaton03@kasi.re.kr}

\author[0000-0003-0562-5643]{Seung-Lee Kim} % ONLY for events with years <= 2023
\affiliation{Korea Astronomy and Space Science Institute, Daejeon 34055, Republic of Korea}
%\email{slkim@kasi.re.kr}

\author[0000-0003-0043-3925]{Chung-Uk Lee}
\affiliation{Korea Astronomy and Space Science Institute, Daejeon 34055, Republic of Korea}
%\email{leecu@kasi.re.kr}

\author[0009-0000-5737-0908]{Dong-Joo Lee} % ONLY for events with years <= 2023
\affiliation{Korea Astronomy and Space Science Institute, Daejeon 34055, Republic of Korea}
%\email{marin678@kasi.re.kr}

\author[0000-0001-7594-8072]{Yongseok Lee} % ONLY for events with years <= 2023
\affiliation{Korea Astronomy and Space Science Institute, Daejeon 34055, Republic of Korea}
\affiliation{School of Space Research, Kyung Hee University, Yongin, Kyeonggi 17104, Republic of Korea}
%\email{yslee@kasi.re.kr}

\author[0000-0002-6982-7722]{Byeong-Gon Park}
\affiliation{Korea Astronomy and Space Science Institute, Daejeon 34055, Republic of Korea}
%\email{bgpark@kasi.re.kr}

\author[0000-0003-1435-3053]{Richard W. Pogge} % ONLY for events with years <= 2023
\affiliation{Department of Astronomy, Ohio State University, 140 West 18th Ave., Columbus, OH  43210, USA}
\affiliation{Center for Cosmology and AstroParticle Physics, Ohio State University, 191 West Woodruff Ave., Columbus, OH 43210, USA}
%\email{pogge.1@osu.edu}

\collaboration{(The KMTNet Collaboration)}

\author{Berto Monard}
\affiliation{Klein Karoo Observatory, Calitzdorp, and Bronberg Observatory, Pretoria, South Africa}
%astroberto13m@gmail.com

\author{Yunyi Tang}
\affiliation{Department of Astronomy, Tsinghua University, Beijing 100084, China}
%tangyy24@mails.tsinghua.edu.cn

\author{Subo Dong}
\affiliation{Department of Astronomy, School of Physics, Peking University, Yiheyuan Rd. 5, Haidian District, Beijing 100871, China}
\affiliation{Kavli Institute for Astronomy and Astrophysics, Peking University, Yiheyuan Rd. 5, Haidian District, Beijing 100871, China}
\affiliation{National Astronomical Observatories, Chinese Academy of Sciences, 20A Datun Road, Chaoyang District, Beijing 100101, China}
%dongsubo@pku.edu.cn

\author{Zhuokai Liu}
\affiliation{Department of Astronomy, School of Physics, Peking University, Yiheyuan Rd. 5, Haidian District, Beijing 100871, China}
\affiliation{Kavli Institute for Astronomy and Astrophysics, Peking University, Yiheyuan Rd. 5, Haidian District, Beijing 100871, China}
%2201110294@stu.pku.edu.cn

\author{Grant Christie}
\affiliation{Auckland Observatory, Auckland, New Zealand}
%grant@christie.org.nz

\author{Jennie McCormick}
\affiliation{Farm Cove Observatory, Centre for Backyard Astrophysics, Pakuranga, Auckland, New Zealand}
%farmcoveobs@xtra.co.nz

\author{Tim Natusch}
\affiliation{Auckland Observatory, Auckland, New Zealand}
\affiliation{Institute for Radio Astronomy and Space Research (IRASR), AUT University, Auckland, New Zealand}
%tim.natusch@aut.ac.nz

\author[0000-0003-4625-8595]{Qiyue Qian}
\affiliation{Department of Astronomy, Tsinghua University, Beijing 100084, China}
%qqy22@mails.tsinghua.edu.cn

\author{Dan Maoz}
\affiliation{School of Physics and Astronomy, Tel-Aviv University, Tel-Aviv 6997801, Israel}
%maoz@astro.tau.ac.il

\author[0000-0003-4027-4711]{Wei Zhu}
\affiliation{Department of Astronomy, Tsinghua University, Beijing 100084, China}
%weizhu@mail.tsinghua.edu.cn

\collaboration{(The MAP \& $\mu$FUN Follow-up Team)}

\author[0000-0001-7016-1692]{Przemek Mr\'{o}z}
\affiliation{Astronomical Observatory, University of Warsaw, Al. Ujazdowskie 4, 00-478 Warszawa, Poland}
%pmroz@caltech.edu

\author[0000-0002-0548-8995]{Micha{\l}~K. Szyma\'{n}ski}
\affiliation{Astronomical Observatory, University of Warsaw, Al. Ujazdowskie 4, 00-478 Warszawa, Poland}
%msz@astrouw.edu.pl

\author[0000-0002-2335-1730]{Jan Skowron}
\affiliation{Astronomical Observatory, University of Warsaw, Al. Ujazdowskie 4, 00-478 Warszawa, Poland}
%jskowron@astrouw.edu.pl

\author[0000-0002-9245-6368]{Radoslaw Poleski}
\affiliation{Astronomical Observatory, University of Warsaw, Al. Ujazdowskie 4, 00-478 Warszawa, Poland}
%rpoleski@astrouw.edu.pl

\author[0000-0002-7777-0842]{Igor Soszy\'{n}ski}
\affiliation{Astronomical Observatory, University of Warsaw, Al. Ujazdowskie 4, 00-478 Warszawa, Poland}
%soszynsk@astrouw.edu.pl

\author[0000-0002-2339-5899]{Pawe{\l} Pietrukowicz}
\affiliation{Astronomical Observatory, University of Warsaw, Al. Ujazdowskie 4, 00-478 Warszawa, Poland}
%pietruk@astrouw.edu.pl

\author[0000-0003-4084-880X]{Szymon Koz{\l}owski}
\affiliation{Astronomical Observatory, University of Warsaw, Al. Ujazdowskie 4, 00-478 Warszawa, Poland}
%simkoz@astrouw.edu.pl

\author[0000-0002-9326-9329]{Krzysztof A. Rybicki}
\affiliation{Astronomical Observatory, University of Warsaw, Al. Ujazdowskie 4, 00-478 Warszawa, Poland}
\affiliation{Department of Particle Physics and Astrophysics, Weizmann Institute of Science, Rehovot 76100, Israel}
%krybicki@astrouw.edu.pl

\author[0000-0002-6212-7221]{Patryk Iwanek}
\affiliation{Astronomical Observatory, University of Warsaw, Al. Ujazdowskie 4, 00-478 Warszawa, Poland}
%piwanek@astrouw.edu.pl

\author[0000-0001-6364-408X]{Krzysztof Ulaczyk}
\affiliation{Department of Physics, University of Warwick, Gibbet Hill Road, Coventry, CV4~7AL,~UK}
%kulaczyk@astrouw.edu.pl

\author[0000-0002-3051-274X]{Marcin Wrona}
\affiliation{Astronomical Observatory, University of Warsaw, Al. Ujazdowskie 4, 00-478 Warszawa, Poland}
\affiliation{Villanova University, Department of Astrophysics and Planetary Sciences, 800 Lancaster Ave., Villanova, PA 19085, USA}
%mwrona@astrouw.edu.pl

\author[0000-0002-1650-1518]{Mariusz Gromadzki}
\affiliation{Astronomical Observatory, University of Warsaw, Al. Ujazdowskie 4, 00-478 Warszawa, Poland}
%Mazki@astrouw.edu.pl

\author{Mateusz J. Mr\'{o}z}
\affiliation{Astronomical Observatory, University of Warsaw, Al. Ujazdowskie 4, 00-478 Warszawa, Poland}

\collaboration{(The OGLE Collaboration)}

\author{Fumio Abe}
\affiliation{Institute for Space-Earth Environmental Research, Nagoya University, Nagoya 464-8601, Japan}
%abe@isee.nagoya-u.ac.jp

\author{Ken Bando}
\affiliation{Department of Earth and Space Science, Graduate School of Science, Osaka University, Toyonaka, Osaka 560-0043, Japan}

\author{David P. Bennett}
\affiliation{Code 667, NASA Goddard Space Flight Center, Greenbelt, MD 20771, USA}
\affiliation{Department of Astronomy, University of Maryland, College Park, MD 20742, USA}
%david.p.bennett@nasa.gov

\author{Aparna Bhattacharya}
\affiliation{Code 667, NASA Goddard Space Flight Center, Greenbelt, MD 20771, USA}
\affiliation{Department of Astronomy, University of Maryland, College Park, MD 20742, USA}
%abhatta5@umd.edu

\author{Ian A. Bond}
\affiliation{Institute of Natural and Mathematical Sciences, Massey University, Auckland 0745, New Zealand}
%I.A.Bond@massey.ac.nz

\author{Akihiko Fukui}
\affiliation{Department of Earth and Planetary Science, Graduate School of Science, The University of Tokyo, 7-3-1 Hongo, Bunkyo-ku, Tokyo 113-0033, Japan}
\affiliation{Instituto de Astrof\'isica de Canarias, V\'ia L\'actea s/n, E-38205 La Laguna, Tenerife, Spain}
%afukui@eps.s.u-tokyo.ac.jp

\author{Ryusei Hamada}
\affiliation{Department of Earth and Space Science, Graduate School of Science, Osaka University, Toyonaka, Osaka 560-0043, Japan}

\author{Shunya Hamada}
\affiliation{Department of Earth and Space Science, Graduate School of Science, Osaka University, Toyonaka, Osaka 560-0043, Japan}

\author{Naoto Hamasak}
\affiliation{Department of Earth and Space Science, Graduate School of Science, Osaka University, Toyonaka, Osaka 560-0043, Japan}

\author{Yuki Hirao}
\affiliation{Department of Earth and Space Science, Graduate School of Science, Osaka University, Toyonaka, Osaka 560-0043, Japan}
%hirao@iral.ess.sci.osaka-u.ac.jp

\author{Stela Ishitani Silva}
\affiliation{Department of Physics, The Catholic University of America, Washington, DC 20064, USA}
\affiliation{Code 667, NASA Goddard Space Flight Center, Greenbelt, MD 20771, USA}
%StelaIS@gmail.com

\author{Naoki Koshimoto}
\affiliation{Department of Astronomy, University of Maryland, College Park, MD 20742, USA}
%koshimoto@astron.s.u-tokyo.ac.jp

\author{Yutaka Matsubara}
\affiliation{Institute for Space-Earth Environmental Research, Nagoya University, Nagoya 464-8601, Japan}
%ymatsu@isee.nagoya-u.ac.jp

\author{Shota Miyazaki}
\affiliation{Department of Earth and Space Science, Graduate School of Science, Osaka University, Toyonaka, Osaka 560-0043, Japan}
%miyazaki@iral.ess.sci.osaka-u.ac.jp

\author{Yasushi Muraki}
\affiliation{Institute for Space-Earth Environmental Research, Nagoya University, Nagoya 464-8601, Japan}
%ysa_muraki@topaz.plala.or.jp

\author{Tutumi Nagai}
\affiliation{Institute for Space-Earth Environmental Research, Nagoya University, Nagoya 464-8601, Japan}

\author{Kansuke Nunota}
\affiliation{Institute for Space-Earth Environmental Research, Nagoya University, Nagoya 464-8601, Japan}

\author{Greg Olmschenk}
\affiliation{Code 667, NASA Goddard Space Flight Center, Greenbelt, MD 20771, USA}
%golmschenk@usra.edu

\author{Cl\'ement Ranc}
\affiliation{Sorbonne Universit\'e, CNRS, Institut d'Astrophysique de Paris, IAP, F-75014, Paris, France}
%clement.ranc@protonmail.com

\author{Nicholas J. Rattenbury}
\affiliation{Department of Physics, University of Auckland, Private Bag 92019, Auckland, New Zealand}
%n.rattenbury@auckland.ac.nz

\author{Yuki Satoh}
\affiliation{Department of Earth and Space Science, Graduate School of Science, Osaka University, Toyonaka, Osaka 560-0043, Japan}
%satohyk@iral.ess.sci.osaka-u.ac.jp

\author{Daisuke Suzuki}
\affiliation{Department of Earth and Space Science, Graduate School of Science, Osaka University, Toyonaka, Osaka 560-0043, Japan}

\author{Sean Terry}
\affiliation{Code 667, NASA Goddard Space Flight Center, Greenbelt, MD 20771, USA}
\affiliation{Department of Astronomy, University of Maryland, College Park, MD 20742, USA}

\author{Paul J. Tristram}
\affiliation{University of Canterbury Mt.\ John Observatory, P.O. Box 56, Lake Tekapo 8770, New Zealand}
%tristram.p@gmail.com

\author{Aikaterini Vandorou}
\affiliation{Code 667, NASA Goddard Space Flight Center, Greenbelt, MD 20771, USA}
\affiliation{Department of Astronomy, University of Maryland, College Park, MD 20742, USA}

\author{Hibiki Yama}
\affiliation{Department of Earth and Space Science, Graduate School of Science, Osaka University, Toyonaka, Osaka 560-0043, Japan}

\collaboration{(The MOA Collaboration)}

\begin{abstract}

We present the discovery and analysis of the sixth microlensing two-planet system, KMT-2022-BLG-1818Lb,c, detected by a follow-up program targeting high-magnification events. Both planets are subject to the well-known ``Close/Wide'' degeneracy, although for the first planet, which has a super-Jovian mass ratio of $q_2 \simeq 5\times 10^{-3}$ in both solutions, the Close topology, with a normalized separation of $s\simeq 0.70$, is clearly preferred by $\Delta\chi^2=26$. However, contrary to all previous two-planet microlensing systems, the mass ratio for the second planet, $q_3$, is substantially (factor of $\sim 10$) different for the Close and Wide topologies of the first planet. While this degeneracy is resolved in the present case due to high-cadence follow-up observations, the appearance of this new degeneracy indicates the need for caution in the analysis of future two-planet systems. A Bayesian analysis suggests that the host is likely a K-dwarf star in the Galactic disk. The first planet is probably a super-Jupiter on a Jupiter-like orbit, while the second planet is a Saturn-class planet on either a Mercury-like or Saturn-like orbit.

\end{abstract}

\section{Introduction}\label{sec:intro}

Multiplanet systems are a common outcome of planet formation and evolution. Among the nearly 6,000 confirmed exoplanets, approximately 40\% reside in systems with two or more planets\footnote{\url{http://exoplanetarchive.ipac.caltech.edu}, as of 2025 April 15.}. The architecture of these systems, particularly the diversity in orbital separations, eccentricities, and inclinations, is widely believed to be shaped by post-formation dynamical interactions, such as planet-planet scattering, secular evolution, and disk migration (e.g., \citealt{Chatterjee2008, Naoz2011, Dawson2018, Carrera2019}). These processes can redistribute planets within a system, leading to the formation of eccentric or inclined orbits, planetary ejections, or the inward migration of planets to short-period orbits.

Gas giant planets, in particular, often play an outsized role in shaping the dynamical architecture of their systems. In the Solar System, for example, Jupiter and Saturn dominate the budgets of both planetary mass and angular momentum. Their orbital evolution is thought to have driven major events in Solar System history, including the clearing and sculpting of the asteroid belt and the Late Heavy Bombardment (e.g., \citealt{Walsh2011, Clement2019}). Moreover, giant planets can act as dynamical barriers, preventing or disrupting the inward migration of smaller planets (e.g., \citealt{Izidoro2015}), or conversely, destabilizing inner systems through resonance crossings or close encounters (e.g., \citealt{Mustill2015}).

The presence of multiple massive planets, particularly cold Jupiters and their siblings, offers valuable insight into the early dynamical history of planetary systems. Recent statistical studies suggest a correlation between cold Jupiters and the occurrence of inner, smaller companions, indicating a possible co-formation or co-evolution scenario (e.g., \citealt{Zhuwu2018, Bryan2019}). Understanding these systems is thus essential for constraining models of planet formation and migration, as well as for placing our own Solar System in a broader Galactic context. For a comprehensive review of the dynamics and demographics of multiplanet systems, we refer readers to \cite{ZhuDong2021}.

The gravitational microlensing technique \citep{Shude1991,Andy1992,BennettRhie} is sensitive to planets on Jupiter-like orbits \citep{Mao2012,Gaudi2012} and has so far led to the discovery of more than 230 exoplanets. However, detecting multiplanet systems through microlensing is challenging for several reasons. First, microlensing is inherently most sensitive to planets near the Einstein ring, limiting the range of detectable separations \citep{Andy1992,OB190960}. Second, to detect multiple planets, the source star must pass close enough to the caustics of each planet to produce detectable perturbations in the light curve. Third, the presence of one planet can suppress the detectability of another, particularly if the second is less massive \citep{kuang2023}. To date, only five unambiguous two-planet systems have been discovered by microlensing: OGLE-2006-BLG-109L \citep{OB06109,OB06109_Dave}, OGLE-2012-BLG-0026L \citep{OB120026,OB120026_AO,OB120026_zhu}, OGLE-2018-BLG-1011 \citep{OB181011}, OGLE-2019-BLG-0468L \citep{OB190468}, and KMT-2021-BLG-1077 \citep{KB211077}.

High-magnification (HM) microlensing events provide an efficient channel for the detection of multiplanet systems. In these events, the source trajectory passes close to the lens star, where the central or resonant caustics associated with each planet are clustered \citep{Gaudi1998_high}. Four of the five previously known microlensing multiplanet systems were discovered in HM events, with maximum magnfications, $A_{\rm max} > 80$, for the underlying single-lens event. For the remaining event, OGLE-2018-BLG-1011, $A_{\rm max} = 27$. Moreover, because the planetary signals in such events are concentrated near the peak and the peak is often bright and predictable, they are well-suited for follow-up observations using small-aperture telescopes ($\lesssim 1~$m).

Here, we report the discovery of the sixth microlensing two-planet system, KMT-2022-BLG-1818L, identified through a targeted follow-up program for high-magnification events alerted by the Korea Microlensing Telescope Network (KMTNet; \citealt{KMT2016}). The follow-up program, led by the Microlensing Astronomy Probe (MAP\footnote{\url{http://i.astro.tsinghua.edu.cn/~smao/MAP/}}) collaboration, utilized the Las Cumbres Observatory Global Telescope (LCOGT) network, supplemented with data from KMTNet (including auto-follow-up data) and the Microlensing Follow-Up Network ($\mu$FUN; \citealt{mufun}). This follow-up program has already played a significant or decisive role in the detection of eight microlensing planets \citep{KB200414,KB210171,KB220440,KB211547,KB220371,KB231431}. Notably, one of these systems, KMT-2020-BLG-0414L, hosts two low-mass companions, an Earth-mass planet and a second object near the brown dwarf/planet boundary \citep{KB200414}, and Keck adaptive optics imaging revealed the host to be a white dwarf \citep{KB200414_AO}.

This paper is organized as follows: In Section~\ref{sec:obser}, we describe the observations and data reduction. Section~\ref{sec:2L} and Section~\ref{sec:3L} present the modeling of the light curve using binary- and triple-lens models, respectively. In Section~\ref{sec:lens}, we derive the physical properties of the lens system. Finally, we discuss the broader implications of this discovery in Section~\ref{sec:dis}.

\section{Observations}\label{sec:obser}

\begin{figure*}
    \centering
   \includegraphics[width=0.85\textwidth]{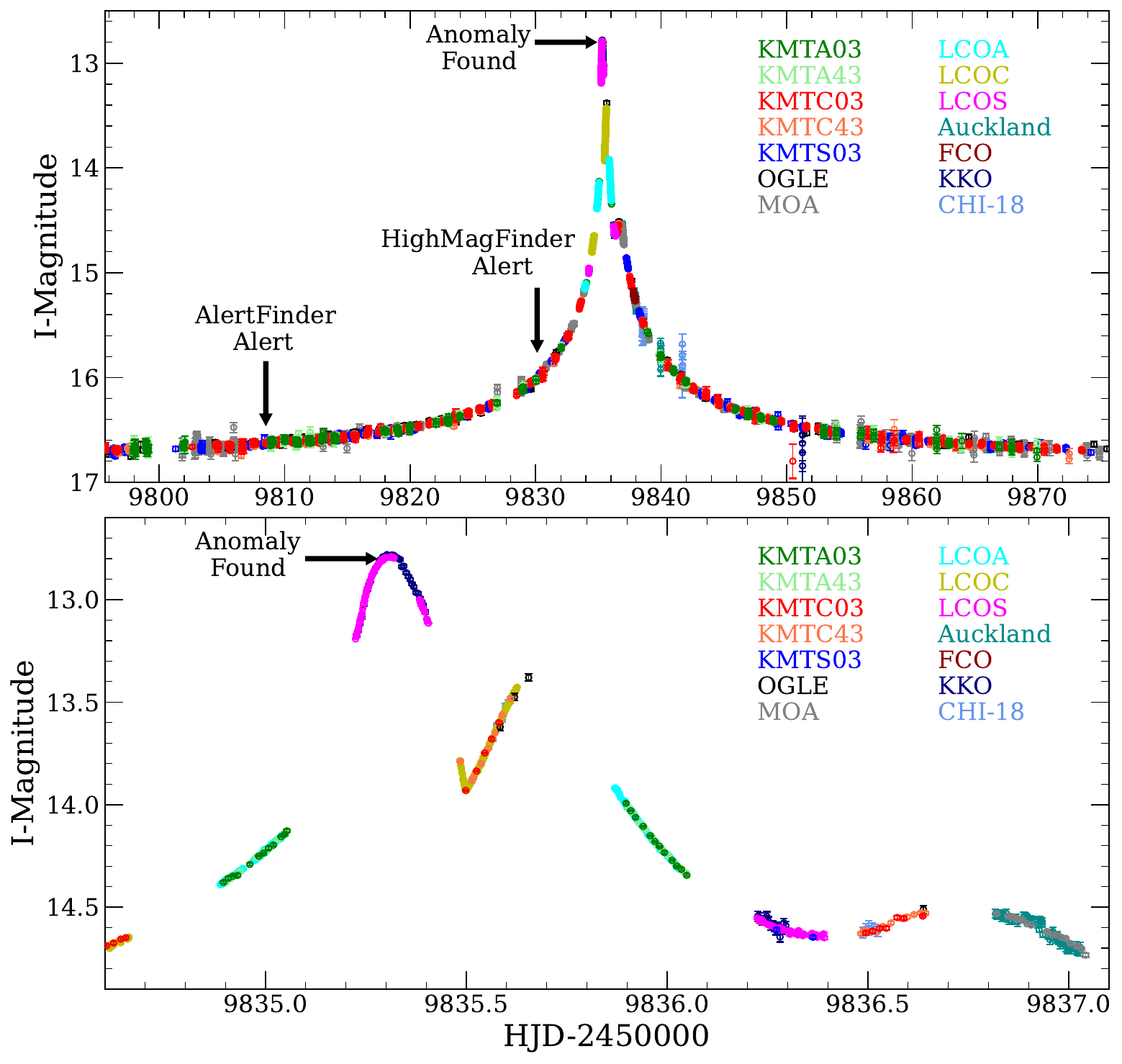}
    \caption{Light curve of the microlensing event, \event. Data point colors correspond to the datasets indicated in the legend. In the upper panel, the three arrows mark (from left to right) the time of discovery by the KMTNet AlertFinder system, the high-magnification alert issued by the KMTNet HighMagFinder system, and the observed anomaly. The lower panel provides a zoomed-in view of the peak region.}
\label{fig:lc1}
\end{figure*}

\subsection{Survey Observations}

Figure~\ref{fig:lc1} presents all light curves collected for this event. The source star of \event\ is located at equatorial coordinates $(\alpha, \delta)_{\rm J2000}$ = (18:03:11.67, $-$27:28:52.00) and Galactic coordinates $(\ell, b) = (3.21, -2.60)$. The event was first alerted by the KMTNet AlertFinder system \citep{KMTAF} on 2022 August 17. KMTNet observations were conducted using its three identical 1.6-meter telescopes located at the Siding Spring Observatory (SSO) in Australia (KMTA), the Cerro Tololo Inter-American Observatory (CTIO) in Chile (KMTC), and the South African Astronomical Observatory (SAAO) in South Africa (KMTS). This event fell within the overlap of two KMTNet fields, BLG03 and BLG43, resulting in a combined cadence of $\Gamma = 4.0~{\rm hr}^{-1}$ for KMTC, and $\Gamma = 6.0~{\rm hr}^{-1}$ for both KMTA and KMTS. For the KMTNet field placement, see Figure 12 of \cite{KMTeventfinder}.

A zoom-in view of the light-curve peak is shown in the lower panel of Figure~\ref{fig:lc1}. Due to scheduled maintenance, KMTS did not observe the first bump. Additionally, KMTC data taken during $9835.63 < \hjd < 9835.66$ were excluded from the analysis due to CCD nonlinearity effects.

The event was independently discovered by the Microlensing Observations in Astrophysics (MOA, \citealt{Sako2008}) collaboration and designated as MOA-2022-BLG-462 \citep{Bond2001}. MOA used the 1.8-meter telescope at Mt.\ John University Observatory in New Zealand, observing the field with a cadence of $\Gamma = 3~{\rm hr}^{-1}$.

Separately, the Optical Gravitational Lensing Experiment (OGLE, \citealt{OGLEIV}) also identified this event as OGLE-2022-BLG-0093 via its post-season Early Warning System \citep{Udalski1994, Udalski2003}. The event was in OGLE field BLG511, which was monitored with a cadence of $\Gamma = 1.0~{\rm hr}^{-1}$ using the 1.3-meter Warsaw Telescope at Las Campanas Observatory in Chile.

Most KMTNet and OGLE images were acquired in the $I$ band, and most MOA images were obtained in the MOA-Red band, which is roughly the sum of the standard Cousins $R$ and $I$ band. The three surveys all took a fraction of $V$-band images for the source color measurements.

\subsection{Follow-up Observations}

At UT 14:54 on 2022 September 7 (${\rm HJD}^{\prime} = 9830.12$), the KMTNet HighMagFinder system \citep{KB210171} issued an alert indicating that this event was expected to reach high magnification within six days. In response, the Microlensing Astronomy Probe (MAP) collaboration initiated follow-up observations using the 1.0-meter Las Cumbres Observatory Global Telescope (LCOGT) network at SSO (LCOA), CTIO (LCOC), and SAAO (LCOS), using the $I$-band filter. The Microlensing Follow-Up Network ($\mu$FUN) also contributed follow-up observations from several sites: the 0.36-meter telescope at Klein Karoo Observatory (KKO) in South Africa, the 0.4-meter telescope at Auckland Observatory (Auckland) in New Zealand, the 0.36-meter telescope at Farm Cove Observatory (FCO) in New Zealand, and a 0.18-meter Newtonian telescope at El Sauce Observatory in Chile (CHI-18).

At ${\rm HJD}^{\prime} = 9835.29$, based on the real-time reduction of LCOS data, W.Zang noticed that this event had deviated from a single-lens model and sent an alert to KMTNet. Then, KMTNet increased the cadence for this event on the subsequent night by replacing the BLG01 observations ($\Gamma = 2~{\rm hr}^{-1}$) with the BLG43 observations.

Additionally, the Observing Microlensing Events of the Galaxy Automatically (OMEGA) Key Project conducted independent follow-up observations of this event. Their results will be presented in a separate paper.

\subsection{Data Reduction}

The difference image analysis (DIA) technique \citep{Tomaney1996,Alard1998} is widely employed in microlensing observations to extract accurate photometry in the crowded stellar fields of the Galactic bulge. For this event, image reductions were performed using custom implementations of DIA for different groups. The KMTNet, LCOGT, and $\mu$FUN data were reduced using the pySIS package \citep{pysis,Yang_TLC}, the MOA data were processed with the pipeline described in \cite{Bond2001}, and the OGLE data were reduced using the DIA algorithm developed by \cite{Wozniak2000}. The error bars from the DIA pipelines were readjusted using the method of \cite{MB11293}, which sets $\chi^2$ per degree of freedom (dof) for each data set to unity. Table~\ref{tab:data} summarizes the observational characteristics and data reduction methods used for each data set in the light-curve modeling.

\begin{table*}[ht]
    \renewcommand\arraystretch{1.1}
    \centering
    \caption{Observation and Reduction Information for Data Used in the Light-curve Analysis}
    \begin{threeparttable}
    \begin{tabular}{c c c c c}
    \hline
    % \hline
    Collaboration & Site & Name & Filter & Reduction Method \\
    \hline
    
    KMTNet & SSO & KMTA03 & $I$ & pySIS\tnote{1} \\

    KMTNet & SSO & KMTA43 & $I$ & pySIS \\
    
    KMTNet & CTIO & KMTC03 & $I$ & pySIS \\

    KMTNet & CTIO & KMTC43 & $I$ & pySIS \\
    
    KMTNet & SAAO & KMTS03 & $I$ & pySIS \\

    KMTNet & SAAO & KMTS43 & $I$ & pySIS \\
    
    OGLE & Las Campanas Observatory & OGLE & $I$ & \cite{Wozniak2000} \\
    
    MOA & Mt. John Observatory & MOA & Red & \cite{Bond2001} \\ 
    
    MAP & SSO & LCOA & $I$ & pySIS \\
    
    MAP & CTIO & LCOC& $I$ & pySIS \\
    
    MAP & SAAO & LCOS & $I$ & pySIS \\
    $\mu$FUN & Klein Karoo Observatory & KKO & unfiltered & pySIS \\
    $\mu$FUN & Auckland Observatory & Auckland & $I$ & pySIS \\ 
    $\mu$FUN & Farm Cove Observatory & FCO & unfiltered & pySIS \\ 
    $\mu$FUN & El Sauce Observatory & CHI-18 & 580--700 nm & pySIS \\
    \hline
    \end{tabular}
   \label{tab:data}
   \begin{tablenotes}
    \item[1] \cite{pysis,Yang_TLC}
    \end{tablenotes}
    \end{threeparttable}
\end{table*}

\section{Double-Lens Analysis}\label{sec:2L}

As shown in Figure \ref{fig:lc1}, the peak of \event\ exhibits three distinct bumps, occurring at approximately $\hjd \sim 9835.3$, 9835.7, and 9836.7. Such a structure cannot be reproduced by a single-lens model, including either a single-lens single-source (1L1S) or single-lens binary-source (1L2S; \citealt{Gaudi1998}) configuration. Therefore, we proceed by fitting the light curve using binary-lens models, specifically considering the binary-lens single-source (2L1S) model and, if necessary, the binary-lens binary-source (2L2S) model.

\begin{table*}
%[htb]
    \renewcommand\arraystretch{1.15}
    \centering
    \caption{Lensing Parameters for 2L1S Models}
    \begin{tabular}{c|c|c|c|c|c|c}
    \hline
    \hline
    \multirow{3}{*}{Parameters} &  \multicolumn{3}{c|}{Close} & \multicolumn{3}{c}{Wide} \\ %\cline{2-7} 
    \hline
     %&
     %$u_0>0$ & $u_0<0$ & $u_0>0$ & $u_0<0$ \\
      & \multicolumn{1}{c|}{Static} & \multicolumn{2}{c|}{Parallax + Orbital Motion} & \multicolumn{1}{c|}{Static} & \multicolumn{2}{c}{Parallax + Orbital Motion}  \\
      & & \multicolumn{1}{c}{ $u_0 > 0$} &  $u_0 < 0$ & & \multicolumn{1}{c}{ $u_0 > 0$} &  $u_0 < 0$ \\
    \hline
    $\chi^2$/dof  
    & $10813.1/10563$ & $10625.2/10559$ & $10621.3/10559$ 
    & $11271.7/10563$ & $10997.8/10559$ & $11013.8/10559$ \\
    \hline
    
    $t_{0} - 9835$ (${\rm HJD}^{\prime}$) 
    & $0.5524 \pm 0.0002$ & $0.5490 \pm 0.0003$ & $0.5485 \pm 0.0003$ 
    & $0.5514 \pm 0.0002$ & $0.5493 \pm 0.0003$ & $0.5486 \pm 0.0003$ \\  
    
    $u_{0} (10^{-3})$ 
    & $3.78 \pm 0.03$ & $3.61 \pm 0.04$ & $-3.58 \pm 0.03$ 
    & $3.79 \pm 0.03$ & $3.54 \pm 0.04$ & $-3.53 \pm 0.04$\\
    
    $\te$ (days) 
    & $72.57 \pm 0.55$ & $74.98 \pm 0.71$ & $75.48 \pm 0.66$ 
    & $72.75 \pm 0.55$ & $76.40 \pm 0.77$ & $76.37 \pm 0.74$ \\
    
    $\rho (10^{-3})$ 
    & $1.48 \pm 0.01$ & $1.44 \pm 0.01$ & $1.43 \pm 0.01$ 
    & $1.47 \pm 0.01$ & $1.41 \pm 0.01$ & $1.41 \pm 0.01$ \\
    
    $\alpha$ (rad) 
    & $2.1846 \pm 0.0005$ & $2.1976 \pm 0.0011$ & $4.0856 \pm 0.0009$ 
    & $2.1895 \pm 0.0005$ & $2.2005 \pm 0.0008$ & $4.0827 \pm 0.0008$ \\
    
    $s$ 
    & $0.7002 \pm 0.0004$ & $0.6936 \pm 0.0006$ & $0.6932 \pm 0.0006$ 
    & $1.4236 \pm 0.0009$ & $1.4310 \pm 0.0011$ & $1.4321 \pm 0.0011$ \\
    
    $q (10^{-3})$ 
    & $4.82 \pm 0.04$ & $4.73 \pm 0.04$ & $4.70 \pm 0.04$ 
    & $4.82 \pm 0.04$ & $4.62 \pm 0.05$ & $4.63 \pm 0.04$\\ 

    $\pi_{\rm E, N}$ 
    & ... & $-0.54 \pm 0.04$ & $0.55 \pm 0.03$ 
    & ... & $-0.76 \pm 0.03$ & $0.76 \pm 0.03$ \\

     $\pi_{\rm E, E}$ 
     & ... & $-0.01 \pm 0.01$ & $-0.03 \pm 0.01$
     & ... & $0.01 \pm 0.01$ & $-0.02 \pm 0.01$ \\

     $ds/dt~({\rm yr}^{-1})$ 
     & ... & $1.58 \pm 0.13$ & $1.67 \pm 0.13$ 
     & ... & $-1.87 \pm 0.11$ & $-1.86 \pm 0.12$  \\

     $d\alpha/dt~(\rm rad~{\rm yr}^{-1})$ 
     & ... & $-1.01 \pm 0.12$ & $0.92 \pm 0.02$ 
     & ... & $0.43 \pm 0.03$ & $-0.43 \pm 0.03$\\

     $\beta$ 
     & ... & $0.83 \pm 0.11$ & $0.84 \pm 0.11$ 
     & ... & $0.98 \pm 0.02$ & $0.98 \pm 0.02$  \\
    
    $I_{\rm S, OGLE}$ 
    & $19.154 \pm 0.009$ & $19.193 \pm 0.011$ & $19.201 \pm 0.010$ 
    & $19.150 \pm 0.009$ & $19.206 \pm 0.012$ & $19.206 \pm 0.011$\\  
    
    \hline
    \hline
    \end{tabular}
    \label{tab:2L}
\end{table*}

\subsection{Binary-lens Single-source}

\begin{figure}
    \centering
    \includegraphics[width=0.47\textwidth]{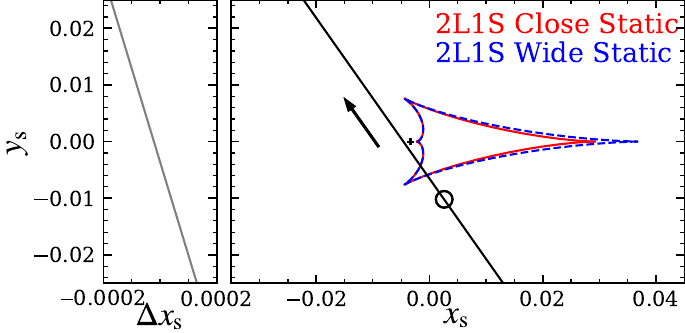}
    \caption{Geometries of the 2L1S Static Models. The right panel shows the caustic structures of the ``Close'' and ``Wide'' models, represented by red and blue curves, respectively. The black cross marks the position of the primary lens, and the black circle indicates the source size. The solid black line denotes the source trajectory, with the arrow indicating the direction of motion for the ``Close'' model. The left panel illustrates the trajectory difference between the ``Wide'' and ``Close'' models, $\Delta x_{\rm S} = x_{\rm S, wide} - x_{\rm S, close}$, shown as a gray line.}
\label{fig:2L1Scau}
\end{figure}

\begin{figure}
    \centering
    \includegraphics[width=0.47\textwidth]{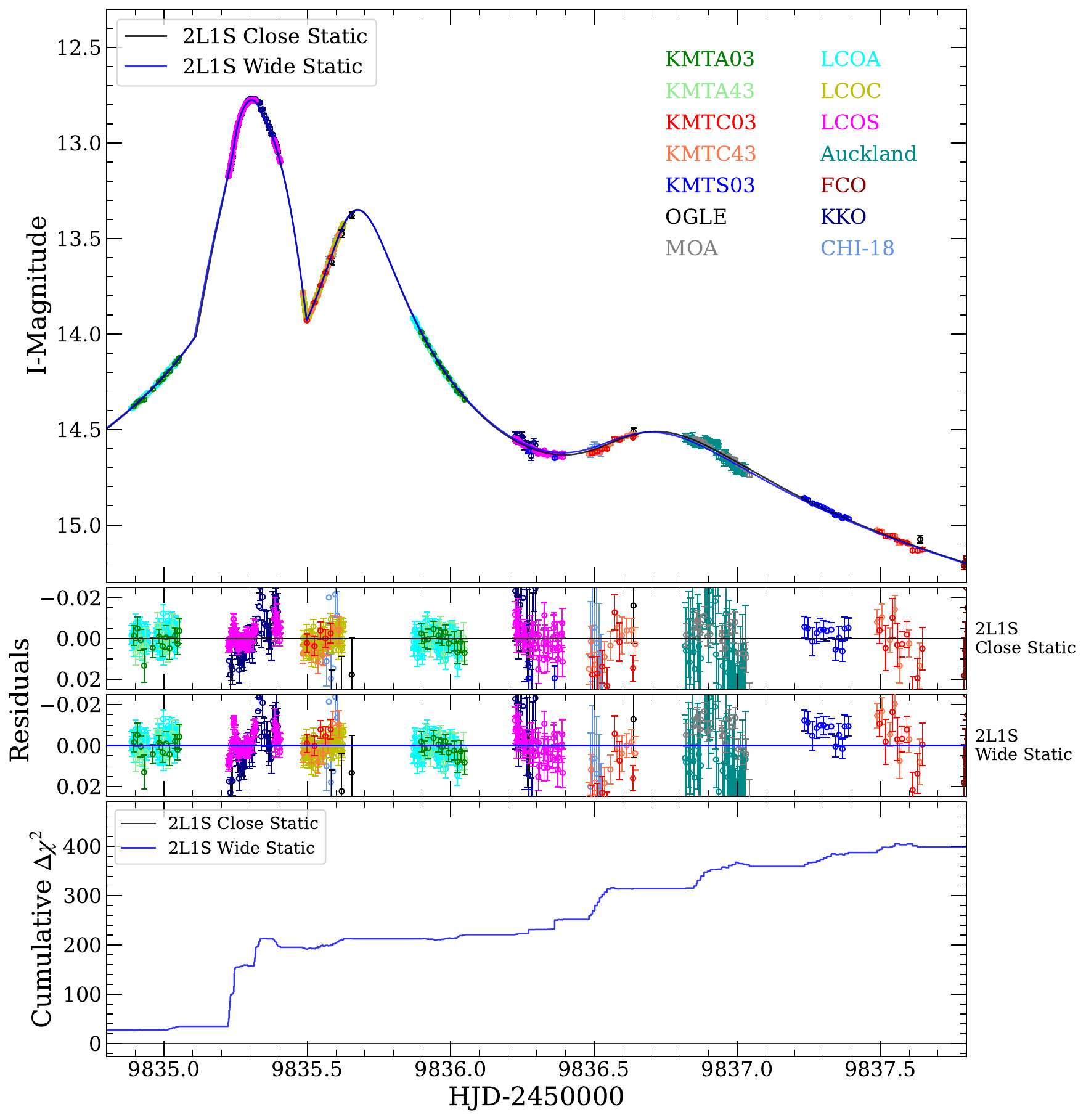}
    \caption{The light curves for the static 2L1S ``Close'' and ``Wide'' model and their residuals to the observed data around the peak. The bottom panel shows the cumulative distribution of $\chi^2$ differences for the ``Wide'' model compared to the ``Close'' model.}
\label{fig:lc2}
\end{figure}

We begin our modeling with the static 2L1S model, which requires seven parameters. Three of these are the standard \citet{Paczynski1986} parameters: $(t_0, u_0, \te)$, representing the time of closest approach between the lens and source, the impact parameter (in units of the angular Einstein radius, $\thetae$), and the Einstein timescale, respectively. The timescale $\te$ and Einstein radius $\thetae$ relate to the lens mass $M_{\rm L}$ by:
\begin{equation}\label{eqn:te}
\te = \frac{\thetae}{\mu_{\rm rel}}; \qquad \thetae = \sqrt{\kappa M_{\rm L} \pi_{\rm rel}},
\end{equation}  
where $\kappa \equiv 4G/(c^2\mathrm{au}) \simeq 8.144~{\rm mas}/{M_{\odot}}$, and $(\pi_{\rm rel}, \mu_{\rm rel})$ are the lens-source relative (parallax, proper motion). Three additional parameters describe the binary-lens geometry: the mass ratio $q$ between the two lens components, their projected separation $s$ (in units of $\thetae$), and the angle $\alpha$ of the source trajectory with respect to the binary axis. The last parameter, $\rho$, represents the ratio of the angular source radius $\theta_*$ to $\thetae$, i.e., $\rho = \theta_*/\thetae$. 

For each dataset $i$, we include two linear flux parameters, $f_{{\rm S},i}$ and $f_{{\rm B},i}$, representing the source flux and any blended flux, respectively. The observed flux at time $t$ is modeled as:
\begin{equation}
    f_{i}(t) = f_{{\rm S},i}A_{\rm 2L1S}(t) + f_{{\rm B},i}. 
\end{equation}
We compute the 2L1S magnification, $A_{\rm 2L1S}(t)$, using the advanced contour integration code \citep{Bozza2010,Bozza2018}, \texttt{VBBinaryLensing}\footnote{\url{http://www.fisica.unisa.it/GravitationAstrophysics/VBBinaryLensing.htm}}. 

Limb-darkening is incorporated using the linear law. Based on the intrinsic color of the source star (see Section \ref{sec:lens}) and the color-temperature relations of \citet{Houdashelt2000}, we estimate an effective temperature of $\sim5570$ K. Assuming $\log g = 2.5$, solar metallicity ([M/H] = 0.0), and a microturbulence of 1~km,s$^{-1}$, we obtain the following limb-darkening coefficients from \citet{Claret2011}: $u_I = 0.51$, $u_V = 0.68$, $u_R = 0.60$, and $u_r = 0.62$. These are assigned to the datasets as follows: $u_I$ for KMTNet, OGLE, LCOGT, and Auckland; $u_V$ for KKO and FCO; $u_r$ for CHI-18; and $u_{\rm MOA} = (u_I + u_R)/2$ for MOA data.

To explore the parameter space, we perform a grid search over $(\log s, \log q, \log \rho, \alpha)$. The grid includes 41 values for $\log s$ uniformly spaced over $[-1.0, 1.0]$, 61 values for $\log q$ over $[-6, 0]$, 6 values for $\log \rho$ over $[-3.5, -2.0]$, and 12 values for $\alpha$ over $[0^\circ, 2\pi)$ rad. For each grid point, we fix $(\log s, \log q)$ and allow $(t_0, u_0, \te, \rho, \alpha)$ to vary, optimizing by Markov chain Monte Carlo (MCMC) using the \texttt{emcee} ensemble sampler \citep{emcee}. This grid search reveals two local minima, with $(\log q, \log s) = (-2.3, -0.15)~{\rm and}~(-2.3, 0.15)$. We refine the two local minima using MCMC, allowing all seven 2L1S parameters to vary. The best-fit models are then obtained via a downhill minimization\footnote{We use a function based on the Nelder–Mead simplex algorithm from the SciPy package. See \url{https://docs.scipy.org/doc/scipy/reference/generated/scipy.optimize.fmin.html\#scipy.optimize.fmin}}. 

Table~\ref{tab:2L} presents the best-fit parameters with their 1$\sigma$ uncertainties, derived from the MCMC posterior distributions. The parameters for the two solutions exhibit the well-known close/wide degeneracy, characterized by approximately symmetric configurations under the transformation $s \leftrightarrow s^{-1}$, with the other parameters almost the same \citep{Griest1998,Dominik1999,An2005}. We refer to these as the ``Close'' ($s < 1$) and ``Wide'' ($s > 1$) solutions. Unlike many other HM events with the close/wide degeneracy, for which the two solutions typically yield comparable $\chi^2$ values, the present event clearly favors the ``Close'' solution, with a $\Delta\chi^2 = 459$. Figure~\ref{fig:2L1Scau} shows the caustic structures and source trajectories for both models. The three observed bumps in the light curve are caused by the source interacting with the central caustic: the first due to a direct crossing, followed by close approaches to the middle-left and upper-left cusps. The caustic shapes and source trajectories of the two solutions exhibit subtle differences. In particular, the source trajectory angles differ by $\Delta\alpha = 0.28^{\circ}$. Thanks to the high-cadence, high-precision coverage of the three bumps, the small differences in the first and third bumps lead to the significant different $\Delta\chi^2$.

However, as shown in the residual panel of Figure~\ref{fig:lc2}, the dense LCOC and LCOS data covering the first two bumps exhibit clear deviations from the two static 2L1S models, with similar trends also in the KMTC and KKO data.

To fit the residuals, we first consider two high-order effects, the microlensing parallax effect \citep{Gould1992,Gould2000,Gouldpies2004} and the lens orbital motion effect \citep{MB09387,OB09020}. The parallax effect arises from deviations in the apparent lens-source rectilinear motion due to Earth's orbital acceleration (i.e., annual parallax) and observations from well-separated ground-based telescopes (i.e., terrestrial parallax, \citealt{OB07224}). This effect is modeled using two parameters, $\pi_{\rm E,N}$ and $\pi_{\rm E,E}$, which represent the north and east components of the microlensing parallax vector $\pie$ in equatorial coordinates, 
\begin{equation}\label{equ:pie}
    \bm{\pi}_{\rm E} = \frac{\pi_{\rm rel}}{\thetae} \frac{\bm{\mu}_{\rm rel}}{\mu_{\rm rel}}.
\end{equation}
To account for the ``ecliptic degeneracy'' \citep{Jiang2004, Poindexter2005}, we fit both the $u_0 > 0$ and $u_0 < 0$ solutions. The lens orbital motion effect is included by two parameters, $ds/dt$ and $d\alpha/dt$, which represent the instantaneous rates of change in the projected separation and orientation of the binary lens components, respectively, evaluated at $t_0$. To ensure that the binary lens system remains bound, we restrict the MCMC trials to $\beta < 1.0$, where $\beta$ is the absolute value of the ratio of transverse kinetic to potential energy \citep{An2002,OB050071D},
\begin{equation}\label{equ:orbital}
    \beta \equiv \left| \frac{{\rm KE}_{\perp}}{{\rm PE}_{\perp}} \right| = \frac{\kappa M_{\odot} {\rm yr}^2}{8\pi^2}\frac{\pie}{\thetae}\eta^2\left(\frac{s}{\pie + \pi_{\rm S}/\thetae}\right)^3,
\end{equation}
\begin{equation}
    \vec{\eta} \equiv \left(\frac{ds/dt}{s}, \frac{d\alpha}{dt}\right),
\end{equation}
where $\pi_{\rm S}$ is the source parallax.

\begin{figure*}
    \centering
    \includegraphics[width=0.48\textwidth]{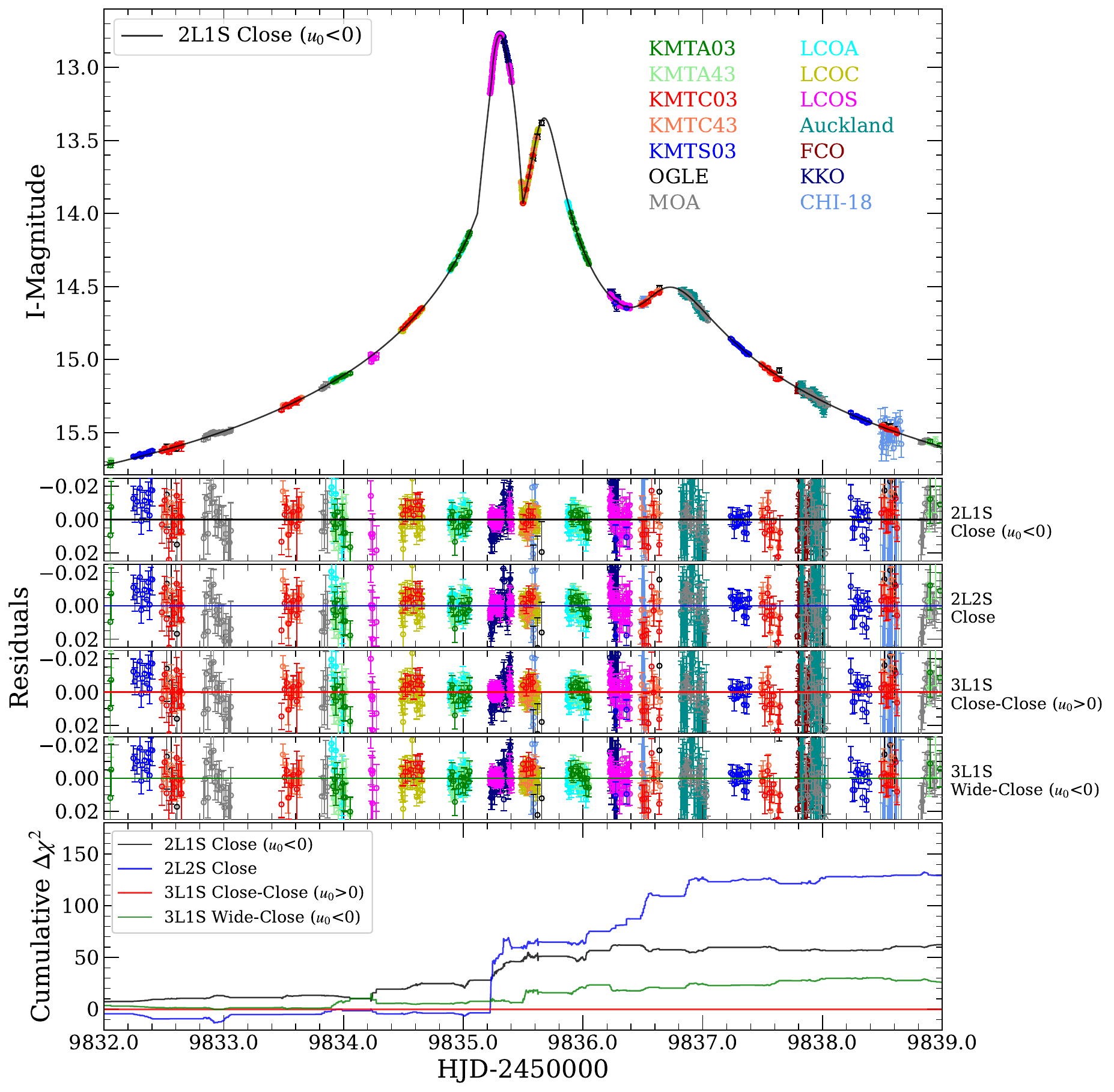}
    \includegraphics[width=0.48\textwidth]{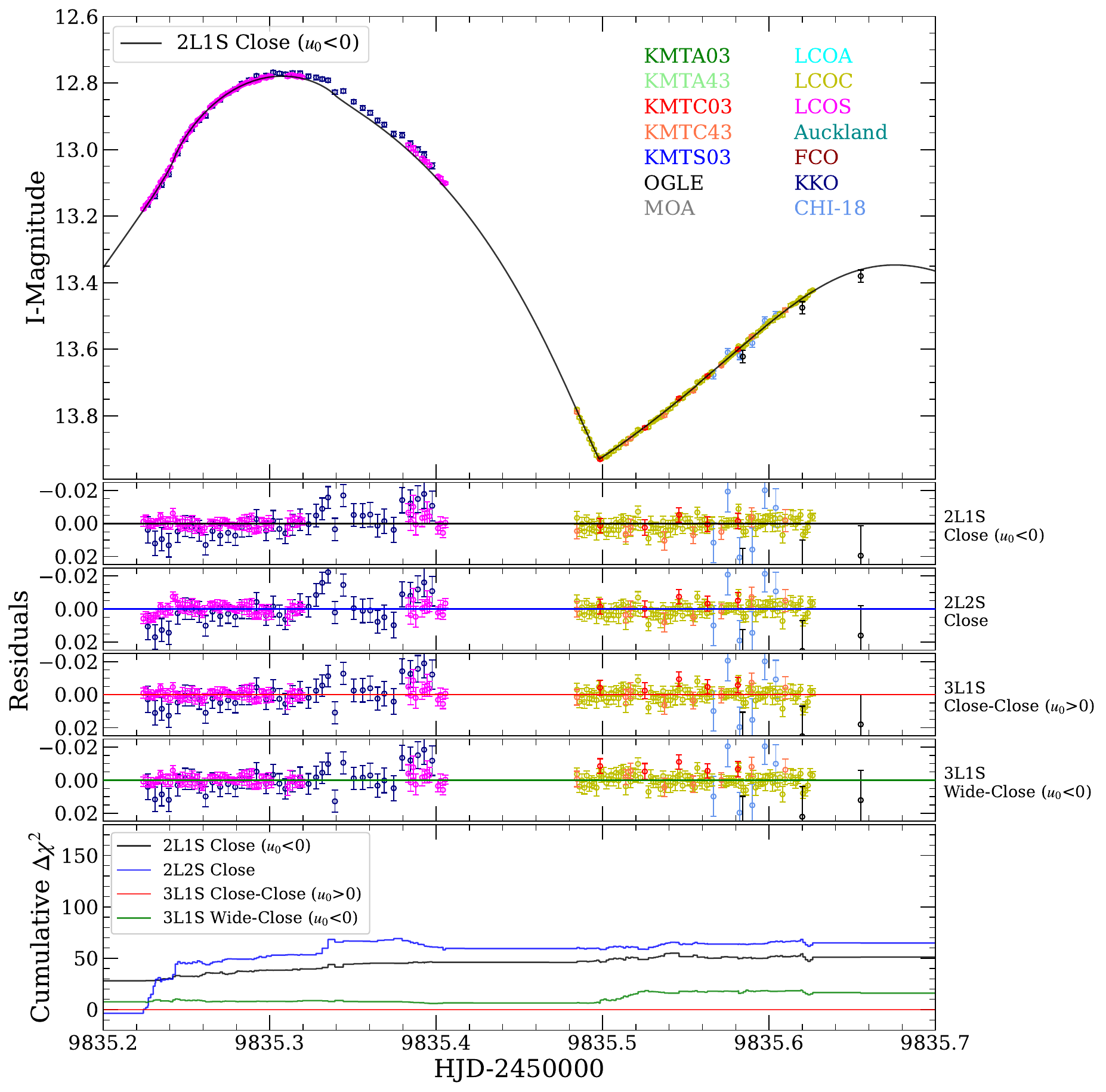}
    \caption{The residuals and cumulative distribution of $\chi^2$ differences over 7 days (left) and 0.5 days (right) centered on the peak, comparing the best-fit 2L1S model with high-order effects and the best-fit models for the 3L1S ``Close'' and ``Wide'' second-body topologies.}
\label{fig:lc3}
\end{figure*}

Table~\ref{tab:2L} summarizes the lensing parameters. Incorporating the two higher-order effects improves the fit by $\Delta\chi^2 = 192$ and $274$ for the Close'' and Wide'' models, respectively. The ``Close'' model remains favored with a $\Delta\chi^2 = 396$. As shown in Figure~\ref{fig:lc3}, the residuals evident in the static 2L1S model are largely accounted for by the inclusion of parallax and lens orbital motion. However, this solution implies a rare, high-eccentricity orbit for the lens system, with $\beta = 0.84 \pm 0.11$. Moreover, the relatively large parallax value, $\pie = 0.55 \pm 0.03$, together with the measured angular Einstein radius $\thetae$ (see Section~\ref{sec:lens}), corresponds to a lens mass of $M \sim 0.15M_{\odot}$ at a distance of approximately 2~kpc. Such a nearby, low-mass lens is uncommon. Therefore, we explore whether the addition of a secondary source to the static 2L1S model can also explain the observed data.

\subsection{Binary-lens Binary-source}

In a 2L2S model, the observed light curve is a flux-weighted combination of two independent 2L1S curves, each corresponding to one of the two source stars. The 2L2S magnification at waveband $\lambda$, $A_{{\rm 2L2S}, \lambda}(t)$, is given by \citep{MB12486}:
\begin{equation}
\begin{aligned}
    A_{{\rm 2L2S}, \lambda}(t) & = \frac{A_{{\rm 2L1S}, 1}(t)f_{1,\lambda} + A_{{\rm 2L1S}, 2}(t)f_{2,\lambda}}{f_{1,\lambda} + f_{2,\lambda}}\\ 
    & = \frac{A_{{\rm 2L1S}, 1}(t) + q_{f,\lambda}A_{{\rm 2L1S}, 2}(t)}{1 + q_{f,\lambda}}, 
\end{aligned}
\end{equation}
where $j = 1, 2$ denotes the primary and secondary sources, respectively, and $f_{j,\lambda}$ represents the source flux in waveband $\lambda$. The flux ratio at waveband $\lambda$ is defined as
\begin{equation}
q_{f,\lambda} = \frac{f_{2,\lambda}}{f_{1,\lambda}}.
\end{equation}

The parameters of the best-fit 2L2S model are presented in Table~\ref{tab:2L2S}. Including the secondary source improves the fit by $\Delta\chi^2 = 109$ and $309$ compared to the 2L1S ``Close'' and ``Wide'' static models, respectively. The 2L2S ``Close'' model remains favored over the ``Wide'' solution by $\Delta\chi^2 = 259$. However, compared to the 2L1S ``Close'' model incorporating higher-order effects, the 2L2S ``Close'' model is disfavored by $\Delta\chi^2 = 83$.

As shown in Figure~\ref{fig:lc3}, the 2L2S ``Close'' model fails to reproduce the LCOS data in the range $9835.25 < \hjd < 9835.28$. Furthermore, according to the analysis in Section~\ref{sec:lens}, the two sources yield inconsistent $\theta_{\rm E}$ values: $\thetae = \theta_{*,1}/\rho_1 \sim 0.7$ mas for the primary source, and $\thetae = \theta_{*,2}/\rho_2 \sim 0.1$ mas for the secondary source. This discrepancy renders the 2L2S ``Close'' model physically implausible.

In conclusion, the 2L2S model fails to adequately describe the observed data. We therefore proceed to investigate whether introducing an additional lens component to the static 2L1S model can explain the light curves.

%% 250507:
\begin{table}
%[htb]
\renewcommand\arraystretch{1.15}
    \centering
    \caption{Lensing Parameters for 2L2S Models}
    \begin{tabular}{c|c|c}
    \hline
    \hline
    Parameters & Close & Wide \\
    \hline
    $\chi^2$/dof & $10704.0/10557$ & $10962.8/10557$ \\
    \hline
    
    $t_{0,1} - 9835$ (${\rm HJD}^{\prime}$) 
    & $0.5524 \pm 0.0004$ & $0.5499 \pm 0.0004$ \\  

    $t_{0,2} - 9835$ (${\rm HJD}^{\prime}$) 
    & $0.6338 \pm 0.1632$ & $0.4520 \pm 0.0026$ \\  
    
    $u_{0,1} (10^{-3})$ 
    & $3.800 \pm 0.006$ & $3.589 \pm 0.037$ \\
    $u_{0,2} (10^{-3})$ 
    & $7.170 \pm 1.897$ & $7.369 \pm 0.104$ \\
    
    $\te$ (days) 
    & $71.92 \pm 0.09$ & $76.27 \pm 0.70$ \\
    
    $\rho_{1} (10^{-3})$ 
    & $1.489 \pm 0.002$ & $1.403 \pm 0.014$ \\
    $\rho_{2} (10^{-3})$ 
    & $1.106 \pm 0.174$ & $0.774 \pm 0.027$ \\
    
    $\alpha$ (rad) 
    & $2.1838 \pm 0.0010$ & $2.1889 \pm 0.0011$ \\
    
    $s$ 
    & $0.6993 \pm 0.0007$ & $1.4210 \pm 0.0013$ \\
    
    $q (10^{-3})$ 
    & $4.889 \pm 0.013$ & $4.600 \pm 0.047$ \\

    $q_{f,\;\rm{I}} (10^{-2})$ 
    & $1.335 \pm 0.844$ & $5.212 \pm 0.688$ \\ 
    $q_{f,\;\rm{V}} (10^{-1})$ 
    & $2.155 \pm 1.597$ & $16.064 \pm 2.903$ \\ 
    $q_{f,\;\rm{R}} (10^{-2})$ 
    & $1.689 \pm 1.127$ & $78.727 \pm 18.839$ \\ 
    
    $I_{\rm S, OGLE}$ 
    & $19.385 \pm 0.002$ & $19.387 \pm 0.023$ \\  

    \hline
    \hline
    \end{tabular}
    \label{tab:2L2S}
\end{table}

\begin{table*}
%[htb]
    \renewcommand\arraystretch{1.15}
    \centering
    \caption{3L1S Parameters for the ``Close'' Topology of the Second Body}
    \begin{tabular}{c|c|c|c|c|c|c}
    \hline
    \hline
    \multirow{3}{*}{Parameters} &  \multicolumn{3}{c|}{Close-Close} & \multicolumn{3}{c}{Close-Wide} \\ %\cline{2-7} 
    \hline
     %&
     %$u_0>0$ & $u_0<0$ & $u_0>0$ & $u_0<0$ \\
      & \multicolumn{1}{c|}{Static} & \multicolumn{2}{c|}{Parallax} & \multicolumn{1}{c|}{Static} & \multicolumn{2}{c}{Parallax}  \\
      & & \multicolumn{1}{c}{ $u_0 > 0$} &  $u_0 < 0$ & & \multicolumn{1}{c}{ $u_0 > 0$} &  $u_0 < 0$ \\
    \hline
    $\chi^2$/dof  
    & $10569.3/10560$ & $10557.7/10558$ & $10562.0/10558$ 
    & $10568.9/10560$ & $10557.9/10558$ & $10562.5/10558$ \\
    \hline
    
    $t_{0} - 9835$ (${\rm HJD}^{\prime}$) 
    & $0.6931 \pm 0.0012$ & $0.6928 \pm 0.0010$ & $0.6922 \pm 0.0011$ 
    & $0.6942 \pm 0.0006$ & $0.6938 \pm 0.0007$ & $0.6936 \pm 0.0009$ \\  
    
    $u_{0} (10^{-3})$ 
    & $0.86 \pm 0.06$ & $0.84 \pm 0.05$ & $-0.81 \pm 0.05$ 
    & $0.91 \pm 0.03$ & $0.90 \pm 0.04$ & $-0.89 \pm 0.05$\\
    
    $\te$ (days) 
    & $72.41 \pm 0.55$ & $73.07 \pm 0.62$ & $73.06 \pm 0.65$ 
    & $72.59 \pm 0.55$ & $73.14 \pm 0.61$ & $73.18 \pm 0.67$ \\
    
    $\rho (10^{-3})$ 
    & $1.51 \pm 0.01$ & $1.50 \pm 0.01$ & $1.50 \pm 0.01$ 
    & $1.51 \pm 0.01$ & $1.50 \pm 0.01$ & $1.50 \pm 0.01$ \\
    
    $\alpha$ (rad) 
    & $2.1802 \pm 0.0013$ & $2.1878 \pm 0.0021$ & $-2.1858 \pm 0.0025$ 
    & $2.1807 \pm 0.0014$ & $2.1874 \pm 0.0032$ & $-2.1847 \pm 0.0033$ \\
    
    $s_2$ 
    & $0.7008 \pm 0.0010$ & $0.7002 \pm 0.0010$ & $0.7003 \pm 0.0010$ 
    & $0.7003 \pm 0.0010$ & $0.6997 \pm 0.0011$ & $0.6998 \pm 0.0012$ \\
    
    $q_2 (10^{-3})$ 
    & $4.85 \pm 0.05$ & $4.82 \pm 0.05$ & $4.82 \pm 0.05$ 
    & $4.85 \pm 0.05$ & $4.82 \pm 0.05$ & $4.82 \pm 0.05$\\

    $s_3$ 
    & $0.2250 \pm 0.0676$ & $0.2010 \pm 0.0506$ & $0.1736 \pm 0.0438$ 
    & $3.3153 \pm 0.6833$ & $3.5399 \pm 0.7805$ & $3.6846 \pm 0.9195$ \\
    
    $q_3 (10^{-3})$ 
    & $1.11 \pm 0.51$ & $1.14 \pm 0.57$ & $1.53 \pm 0.77$ 
    & $0.49 \pm 0.23$ & $0.51 \pm 0.25$ & $0.57 \pm 0.28$\\ 

    $\psi$ (rad)
    & $0.92 \pm 0.08$ & $0.90 \pm 0.08$ & $-0.89 \pm 0.09$
    & $0.96 \pm 0.07$ & $0.95 \pm 0.09$ & $-0.96 \pm 0.08$\\  

    $\pi_{\rm E, N}$ 
    & ... & $-0.34 \pm 0.07$ & $0.26 \pm 0.09$
    & ... & $-0.30 \pm 0.13$ & $0.18 \pm 0.17$ \\

     $\pi_{\rm E, E}$ 
     & ... & $0.00 \pm 0.01$ & $-0.01 \pm 0.01$
     & ... & $-0.00 \pm 0.01$ & $-0.01 \pm 0.01$ \\

     %$\hat{\bm{\pi}}_{\rm E}$ (rad)
     %& ... & $-0.028$&$-0.015$
     %& ... & $-0.019$&$-0.022$\\
    
    $I_{\rm S, OGLE}$ 
    & $19.150 \pm 0.009$ & $19.160 \pm 0.010$ & $19.160 \pm 0.010$ 
    & $19.153 \pm 0.009$ & $19.161 \pm 0.009$ & $19.162 \pm 0.010$\\ 
    
    \hline
    \hline
    \end{tabular}
    \label{tab:3L1SClose}
\end{table*}

\begin{table*}
%[htb]
    \renewcommand\arraystretch{1.15}
    \centering
    \caption{3L1S Parameters for the ``Wide'' Topology of the Second Body}
    \begin{tabular}{c|c|c|c|c|c|c}
    \hline
    \hline
    \multirow{3}{*}{Parameters} &  \multicolumn{3}{c|}{Wide-Close} & \multicolumn{3}{c}{Wide-Wide} \\ %\cline{2-7} 
    \hline
     %&
     %$u_0>0$ & $u_0<0$ & $u_0>0$ & $u_0<0$ \\
      & \multicolumn{1}{c|}{Static} & \multicolumn{2}{c|}{Parallax} & \multicolumn{1}{c|}{Static} & \multicolumn{2}{c}{Parallax}  \\
      & & \multicolumn{1}{c}{ $u_0 > 0$} &  $u_0 < 0$ & & \multicolumn{1}{c}{ $u_0 > 0$} &  $u_0 < 0$ \\
    \hline
    $\chi^2$/dof  
    & $10584.6/10560$ & $10584.1/10558$ & $10583.7/10558$ 
    & $10585.3/10560$ & $10584.8/10558$ & $10584.5/10558$ \\
    \hline
    
    $t_{0} - 9835$ (${\rm HJD}^{\prime}$) 
    & $0.6749 \pm 0.0037$ & $0.6744 \pm 0.0032$ & $0.6745 \pm 0.0034$ 
    & $0.6774 \pm 0.0025$ & $0.6775 \pm 0.0024$ & $0.6765 \pm 0.0025$ \\  
    
    $u_{0} (10^{-3})$ 
    & $0.22 \pm 0.11$ & $0.20 \pm 0.09$ & $-0.20 \pm 0.10$ 
    & $0.30 \pm 0.07$ & $0.30 \pm 0.07$ & $-0.27 \pm 0.07$\\
    
    $\te$ (days) 
    & $71.66 \pm 0.55$ & $71.83 \pm 0.60$ & $71.66 \pm 0.58$ 
    & $71.99 \pm 0.56$ & $72.14 \pm 0.63$ & $71.96 \pm 0.58$ \\
    
    $\rho (10^{-3})$ 
    & $1.54 \pm 0.01$ & $1.54 \pm 0.01$ & $1.54 \pm 0.01$ 
    & $1.54 \pm 0.01$ & $1.53 \pm 0.01$ & $1.54 \pm 0.01$ \\
    
    $\alpha$ (rad) 
    & $2.1755 \pm 0.0013$ & $2.1745 \pm 0.0044$ & $-2.1744 \pm 0.0028$ 
    & $2.1757 \pm 0.0012$ & $2.1743 \pm 0.0042$ & $-2.1745 \pm 0.0023$ \\
    
    $s_2$ 
    & $1.4272 \pm 0.0018$ & $1.4269 \pm 0.0020$ & $1.4263 \pm 0.0017$ 
    & $1.4220 \pm 0.0023$ & $1.4213 \pm 0.0027$ & $1.4212 \pm 0.0023$ \\
    
    $q_2 (10^{-3})$ 
    & $5.06 \pm 0.04$ & $5.05 \pm 0.05$ & $5.05 \pm 0.05$ 
    & $5.02 \pm 0.04$ & $5.00 \pm 0.05$ & $5.02 \pm 0.05$\\

    $s_3$ 
    & $0.1021 \pm 0.0141$ & $0.0996 \pm 0.0124$ & $0.0974 \pm 0.0115$ 
    & $8.9091 \pm 0.8772$ & $8.9973 \pm 0.9747$ & $9.1510 \pm 0.8440$ \\
    
    $q_3 (10^{-3})$ 
    & $9.76 \pm 2.64$ & $10.09 \pm 2.26$ & $10.43 \pm 2.40$ 
    & $7.80 \pm 1.37$ & $7.91 \pm 1.56$ & $8.31 \pm 1.42$\\ 

    $\psi$ (rad)
    & $1.06 \pm 0.03$ & $1.06 \pm 0.03$ & $-1.05 \pm 0.03$
    & $1.08 \pm 0.03$ & $1.07 \pm 0.02$ & $-1.07 \pm 0.02$ \\  

    $\pi_{\rm E, N}$ 
    & ... & $0.04 \pm 0.19$ & $-0.07 \pm 0.10$
    & ... & $0.07 \pm 0.19$ & $-0.05 \pm 0.10$ \\

     $\pi_{\rm E, E}$ 
     & ... & $-0.00 \pm 0.01$ & $-0.00 \pm 0.01$ 
     & ... & $-0.00 \pm 0.01$ & $0.00 \pm 0.01$ \\

     %$\hat{\bm{\pi}}_{\rm E}$ (rad) 
     %& ... &$-0.025$  &$-0.015$
     %& ... &$0.093$  &$-0.0066$\\
    
    $I_{\rm S, OGLE}$ 
    & $19.130 \pm 0.009$ & $19.133 \pm 0.010$ & $19.131 \pm 0.009$ 
    & $19.131 \pm 0.009$ & $19.134 \pm 0.010$ & $19.130 \pm 0.009$\\ 
    
    \hline
    \hline
    \end{tabular}
    \label{tab:3L1SWide}
\end{table*}

\section{Triple-lens Analysis}\label{sec:3L}

In this section, we fit the light curves using the triple-lens single-source (3L1S) model. Compared to the 2L1S model, the 3L1S model introduces three additional parameters to describe the third body: $(s_3, q_3, \psi)$. These correspond to the projected separation between the third body and the primary (scaled to the Einstein radius), the mass ratio between the third body and the primary, and the orientation angle of the third body with respect to the axis connecting the primary and secondary lenses. To avoid confusion, we denote the separation and mass ratio between the secondary and primary as $(s_2, q_2)$ for the 3L1S model.

We begin the 3L1S modeling with a grid search over the parameter space of $(\log s_3, \log q_3, \psi)$. The grid consists of 61 values for $\log s_3$ uniformly spaced over the range $[-1.5, 1.5]$, 61 values for $\log q_3$ over $[-6, 0]$, and 180 values for $\psi$ over $[0, 2\pi)$ radians. Because the ``Close'' model is strongly favored in the 2L1S analysis, we fix the secondary lens parameters, $(s_2, q_2, \alpha)$, to those of the 2L1S ``Close'' model, while allowing $(t_0, u_0, \te, \rho)$ to vary during the MCMC. For each grid point, which defines a fixed lens configuration, we compute the 3L1S magnifications using the map-making technique of \citet{OB04343,MB07400}. The grid search reveals two local minima at $(\log s_3, \log q_3, \psi) = (-0.65, -3.0, 0.91~\mathrm{rad})$ and $(0.50, -3.3, 0.94~\mathrm{rad})$, respectively. We then refine these solutions using MCMC, allowing all 3L1S parameters to vary. Because the lens geometry evolves during the MCMC sampling, we compute the magnifications applying the recently developed contour integration code \texttt{VBMicroLensing} \citep{Bozza2025}\footnote{\url{https://github.com/valboz/VBMicroLensing}}.

The resulting parameters are presented in Table~\ref{tab:3L1SClose}. To maintain consistency with the 2L1S model terminology, we classify the third body as ``Close'' for $s_3 < 1$ and ``Wide'' for $s_3 > 1$. Accordingly, the two 3L1S solutions are labeled as ``Close-Close'' ($s_2 < 1$, $s_3 < 1$) and ``Close-Wide'' ($s_2 < 1$, $s_3 > 1$). The $\chi^2$ difference between the solution is only 0.4. The parameters associated with the secondary body, $s_2$ and $q_2$, are consistent with those of the 2L1S ``Close'' model within 1$\sigma$. Moreover, the projected separations of the third body in the two solutions satisfy the $s \leftrightarrow s^{-1}$ relation within 1$\sigma$, indicative of the close/wide degeneracy. The mass ratios of the third body, $q_3$, are also mutually consistent within 1$\sigma$. We then incorporate the microlensing parallax effect into the 3L1S model. Due to the ecliptic degeneracy, there are $2 \times 2 = 4$ solutions. The inclusion of the parallax effect yields slight improvements in the fit, with $\Delta\chi^2 \sim 11$ for the $u_0 > 0$ solutions and $\sim 6$ for the $u_0 < 0$ solutions.

Compared to the 2L1S model with high-order effects, the 3L1S parallax model is favored by $\Delta\chi^2 = 63.4$. In Figure~\ref{fig:lc3}, we show the residuals for both models, as well as the cumulative distribution of their $\chi^2$ differences. The 3L1S parallax model provides a better fit to the first and second bumps, with an improvement of $\Delta\chi^2 = 24$. Data within $\pm 1$ day of $t_0$ contribute $\Delta\chi^2 = 50$. Moreover, the 3L1S parallax model is also favored in the wings of the light curve, yielding a $\Delta\chi^2 = 14$. It suggests that the large parallax amplitude and the 0.04 magnitude fainter source required by the 2L1S model with high-order effects are disfavored in the light-curve wings. 

We will further compare the two models in Section~\ref{sec:lens}, incorporating the physical parameters of the lens system.

Following the same procedure, we search for 3L1S model based on the static 2L1S ``Wide'' model and identify two additional solutions. We designate these as ``Wide-Close'' ($s_2 > 1$, $s_3 < 1$) and ``Wide-Wide'' ($s_2 > 1$, $s_3 > 1$). The corresponding MCMC and downhill minimization results are presented in Table~\ref{tab:3L1SWide}. As in the 2L1S ``Close'' topology, the third-body parameters, $s_3$ and $q_3$, also exhibit the close/wide degeneracy within 1$\sigma$. 

\begin{figure}
    \centering
    \includegraphics[width=0.47\textwidth]{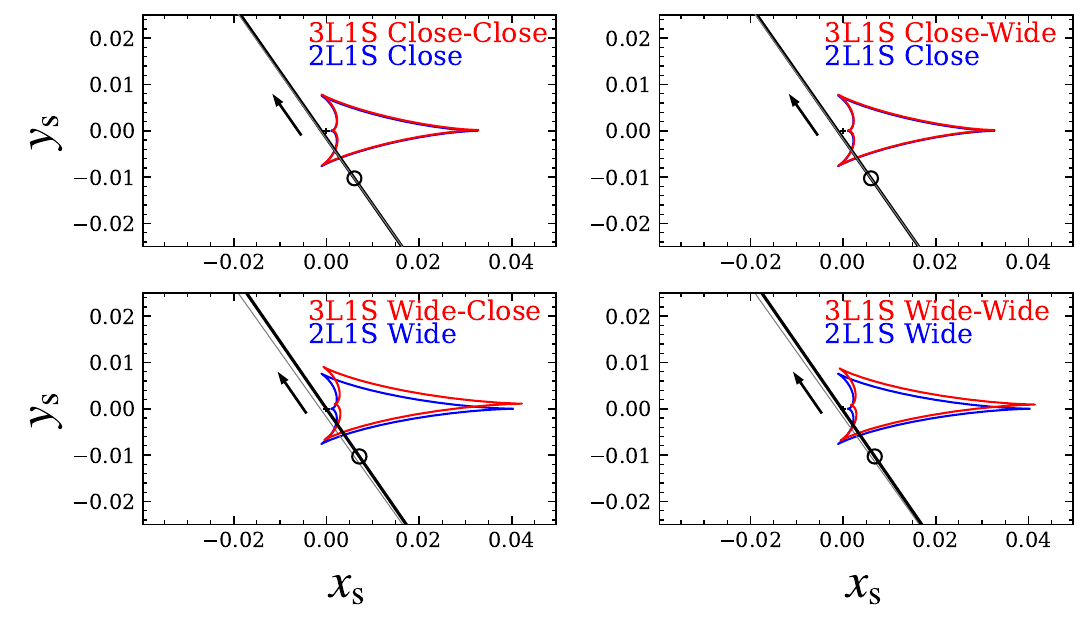}
    \caption{Geometries of the static 3L1S and 2L1S models. The black and gray lines represent the source trajectories for the 3L1S model and its corresponding 2L1S model, respectively. For the ``Close'' second-body topology, the differences between the two trajectories are less significant.}
\label{fig:3L1Scau}
\end{figure}

Although the ``Wide'' topology was strongly disfavored in the 2L1S modeling by $\Delta\chi^2 = 459$, in the 3L1S modeling, it is only slightly disfavored, with a $\Delta\chi^2 = 16$. The parameters of the second body ($s_2$, $q_2$) differ by $\geq 5\sigma$ from those of the 2L1S models. The third-body parameters ($s_3$, $q_3$) vary significantly between the ``Close'' and ``Wide'' second-body topologies, with an order-of-magnitude difference in $q_3$. The caustic geometries are also slightly different between the ``Close'' and ``Wide'' second-body topologies, as shown in Figure~\ref{fig:3L1Scau}. We will discuss these features in Section \ref{sec:32}. 

Across all four 3L1S solutions, the mass ratios of both companions lie within the planetary regime, indicating two planets in the lens system.

In addition, unlike previously known two-planet systems, for which the central caustic is nearly the superposition of two caustics produced by the two planets, respectively, becoming intertwined where they overlap, the third body in this event induces only a subtle perturbation to the shape and position of the central caustic of the second body. Prior to our event, a similar feature in the central caustic was observed in the two-planet model of OGLE-2007-BLG-0349, although subsequent HST data favored a circumbinary interpretation \citep{OB07349}. Our event reveals a new detection channel for multi-planet systems, which we discuss in detail in Section~\ref{sec:31}. 

Incorporating the parallax effect results in a marginal improvement, with $\Delta\chi^2 < 1$, so the ``Close'' second-body topology with parallax is favored by $\Delta\chi^2 = 26.4$. According to Figure~\ref{fig:lc3}, most of the $\chi^2$ difference between the ``Close'' and ``Wide'' topologies arises in the anomalous region of the light curve. This is reasonable, as the differences between the two topologies are expected to appear during the anomaly, and the light curve is particularly sensitive to the parallax effect during the caustic crossing and cusp approach (e.g., \citealt{KB200414}). Given that $\Delta\chi^2 = 26.4$ is statistically significant, we exclude the ``Wide'' topology and adopt the ``Close'' second-body models as our final interpretation.

\section{Source and Lens Properties}\label{sec:lens}

\begin{figure}
    \centering
    \includegraphics[width=0.47\textwidth]{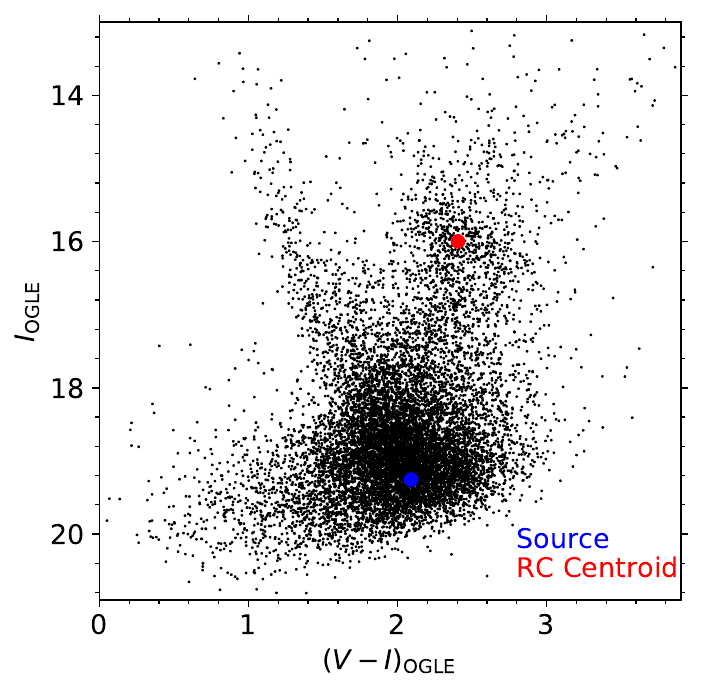}
    \caption{Color–magnitude diagram for field stars within $150^{\prime\prime}$ of \event, constructed from the OGLE-III star catalog \citep{OGLEIII}. The red asterisk marks the centroid of the red giant clump, and the blue dot indicates the position of the microlensing source.}
\label{fig:cmd}
\end{figure}

\begin{table*}
    \renewcommand\arraystretch{1.5}
    \centering
    \caption{Physical Parameters from a Bayesian Analysis and the $\Delta\chi^2$ from different weights}
    \begin{tabular}{c|c c c c c c c c c c}
    \hline
    \hline
     & \multicolumn{7}{c}{Physical Parameters} & \multicolumn{3}{c}{$\Delta\chi^2$} \\
     
    & $M_1$ & $M_2$ & $M_3$ & $D_{\rm L}$ & $a_{\bot, 2}$ & $a_{\bot,3}$ & $P_{\rm disk}$ & Gal.Mod. & Light Curve & Total \\

    Units & $M_{\odot}$ & $M_J$ & $M_J$ & kpc & au & au & & & & \\
    \hline
    Close-Close ($u_0>0$) &$0.30^{+0.12}_{-0.06}$
 & $1.50^{+0.62}_{-0.31}$&$0.35^{+0.24}_{-0.17}$ &$2.9^{+0.7}_{-0.4}$ &$1.5^{+0.3}_{-0.2}$
 &$0.4^{+0.1}_{-0.1}$ &$0.97$ &6.1 & 0.0 & 2.6 \\
    Close-Close ($u_0<0$) &$0.56^{+0.37}_{-0.20}$
 &$2.86^{+1.87}_{-1.02}$ &$0.87^{+0.81}_{-0.49}$ &$4.3^{+1.5}_{-0.9}$
 & $2.1^{+0.7}_{-0.5}$&$0.5^{+0.2}_{-0.2}$
 &$0.88$ &1.0 & 4.3 & 1.8 \\
    Close-Wide ($u_0>0$) &$0.78^{+0.26}_{-0.38}$ &$3.93^{+1.31}_{-1.93}$ &$0.35^{+0.32}_{-0.20}$ &$5.2^{+2.0}_{-1.8}$
 &$2.6^{+1.0}_{-0.9}$ &$12.3^{+6.5}_{-4.5}$
 &$0.59$ &3.3 & 0.2 & 0.0 \\
    Close-Wide ($u_0<0$) &$0.85^{+0.23}_{-0.34}$
 &$4.27^{+1.17}_{-1.70}$ &$0.46^{+0.32}_{-0.25}$
 &$5.3^{+1.8}_{-1.2}$
 & $2.6^{+0.8}_{-0.6}$
&$13.3^{+5.6}_{-4.2}$
 &$0.74$ &0.0 & 4.8 & 1.3\\
    \hline
    \hline
    \end{tabular}
    \tablecomments{ 
Gal.Mod. represents the relative probability from the Galactic model, for which the $\Delta\chi^2$ is derived by $-2\ln({\rm Gal.Mod.})$. The $\Delta\chi^2$ of light-curve analysis are from Tables \ref{tab:3L1SClose}. $P_{\rm disk}$ is the probability of a lens in the Galactic disk.}
\label{tab:phy}
\end{table*}

\subsection{Color-Magnitude Diagram}

To derive the intrinsic properties of the source star, we place it on a color-magnitude diagram (CMD). The CMD is constructed using stars from the OGLE-III catalog \citep{OGLEIII} located within $150''$ of the event location. Following the method of \citet{Nataf2013}, we determine the centroid of the red giant clump as $(V - I, I){\rm cl} = (2.406 \pm 0.011, 15.998 \pm 0.035)$. We adopt the de-reddened color and magnitude of the red giant clump as $(V - I, I){\rm cl,0} = (1.06 \pm 0.03, 14.35 \pm 0.04)$ from \citet{Bensby2013} and Table 1 of \citet{Nataf2013}. These yield an extinction and reddening toward the event direction of $A_I = 1.65 \pm 0.05$ and $E(V - I) = 1.35 \pm 0.03$, respectively.

We estimate the source apparent color as $(V - I)_{\rm S} = 2.092 \pm 0.005$ based on a regression of KMTC03 $V$ versus $I$ flux as a function of magnification, followed by a calibration to OGLE-III magnitude. Because the source color measurement is model-independent, the intrinsic source color is $(V - I)_{\rm S,0} = 0.74 \pm 0.03$. The apparent source magnitude depends on the model. Because the source apparent magnitude depends on the model, for simplicity, we derive results for the 3L1S ``Close-Close'' $u_0 > 0$ model and then present a scaling relation for different source magnitudes. With $I_{\rm S} = 19.160 \pm 0.010$, the intrinsic source magnitude is $I_{\rm S, 0} = 17.51 \pm 0.05$. Applying the color/surface-brightness relation of \cite{Adams2018}, we estimate the angular source radius as 
\begin{equation}
    \theta_* = 1.048 \pm 0.051~\mu {\rm as}.
\end{equation}
This yields an angular Einstein radius of
\begin{equation}
    \thetae = \frac{\theta_*}{\rho} = 0.699 \pm 0.034~\text{mas},
\end{equation}
and a lens-source relative proper motion of
\begin{equation}
    \mu_{\rm rel} = \frac{\thetae}{\te} = 3.49 \pm 0.17~\text{mas yr}^{-1}.
\end{equation}
For any particular model with source magnitude $I_{\rm S}$, the angular source radius can be scaled from the fiducial estimate using $\theta_* = 1.048 \times 10^{-0.2(I_{\rm S} - 19.160)}~\mu {\rm as}$.

\subsection{Bayesian Analysis}

Combining Equations (\ref{eqn:te}) and (\ref{equ:pie}), the lens mass $M_{\rm L}$ and distance $D_{\rm L}$ are related to the angular Einstein radius and microlensing parallax by \citep{Gould1992, Gould2000} 
\begin{equation}\label{eq:mass}
    M_{\rm L} = \frac{\thetae}{{\kappa}\pie};\qquad D_{\rm L} = \frac{\mathrm{au}}{\pie\thetae + \pi_{\rm S}}.
\end{equation}
Due to the relatively large uncertainties in $\pie$ for the 3L1S parallax solutions, we perform a Bayesian analysis by incorporating priors from a Galactic model to estimate the physical parameters of the lens system.

The Galactic model has three key components: the lens mass function, the stellar number density profile, and the dynamical distributions. For the mass function, we adopt the initial mass function of \citet{Kroupa2001}, applying upper mass cutoffs of $1.3M_{\odot}$ and $1.1M_{\odot}$ for disk and bulge lenses, respectively \citep{Zhu2017spitzer}. For the stellar number density profiles, we choose the models used by \cite{Yang2021_GalacticModel}. For the disk velocity distribution, we adopt the dynamically self-consistent ``Model C'' from \citet{Yang2021_GalacticModel}. For the bulge, we follow the kinematic prescription of \citet{Zhu2017spitzer}, assuming zero mean velocity and a velocity dispersion of $120~{\rm km~s^{-1}}$ in each direction.

We create a sample of $10^8$ simulated events drawn from the Galactic model. For each simulated event, $i$, with parameters $\theta_{{\rm E},i}$, $\mu_{{\rm rel},i}$,$t_{{\rm E},i}$, and $\bm{\pi_{\rm E,i}}$, we weight it by
\begin{equation}
    \omega_{{\rm Gal},i} = \theta_{{\rm E},i} \times \mu_{{\rm rel},i} \times \mathcal{L}(t_{{\rm E},i}) \mathcal{L}(\theta_{{\rm E},i}) \mathcal{L}(\bm{\pi_{\rm E,i}}). 
\end{equation}
where $\mathcal{L}(t_{{\rm E},i})$, $\mathcal{L}(\theta_{{\rm E},i})$, and $\mathcal{L}(\bm{\pi_{\rm E,i}})$ are the likelihoods of $t_{{\rm E},i}$, $\theta_{{\rm E},i}$, and $\bm{\pi_{\rm E,i}}$ given the error distributions derived from the MCMC chain of the light-curve analysis. In this analysis, we consider only the primary lens (i.e., the host star), so the simulated values of $t_{{\rm E},i}$ and $\theta_{{\rm E},i}$ are scaled by $1/\sqrt{1 + q_2 + q_3}$.

\begin{figure*}
    \centering
    \includegraphics[width=0.8\textwidth]{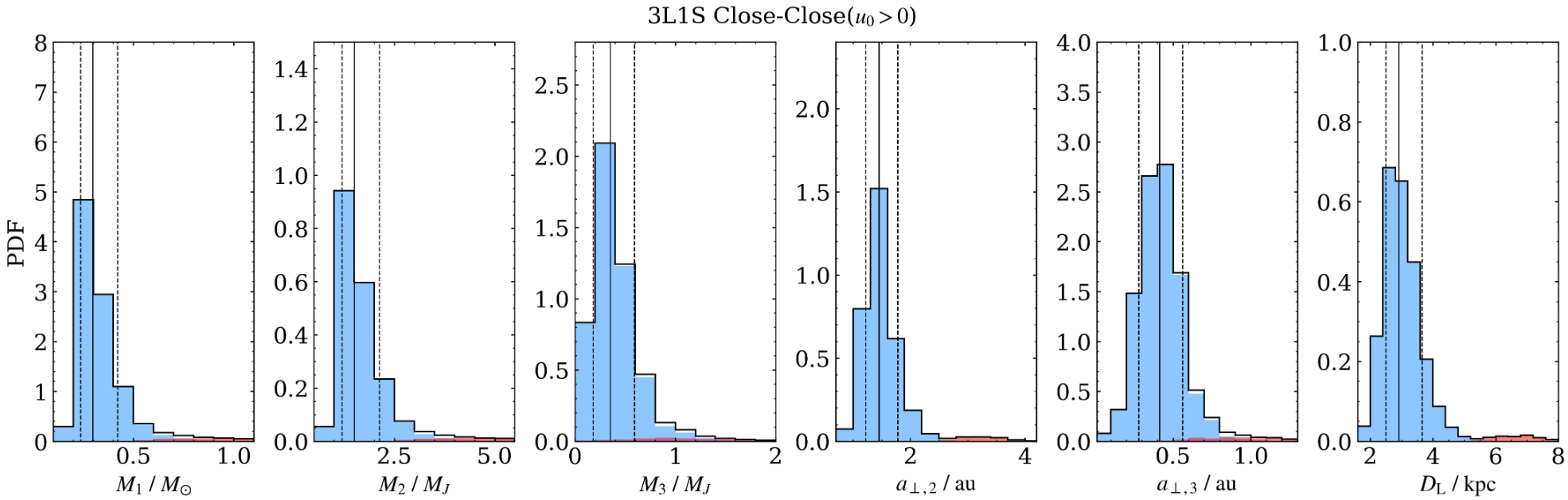}
    \includegraphics[width=0.8\textwidth]{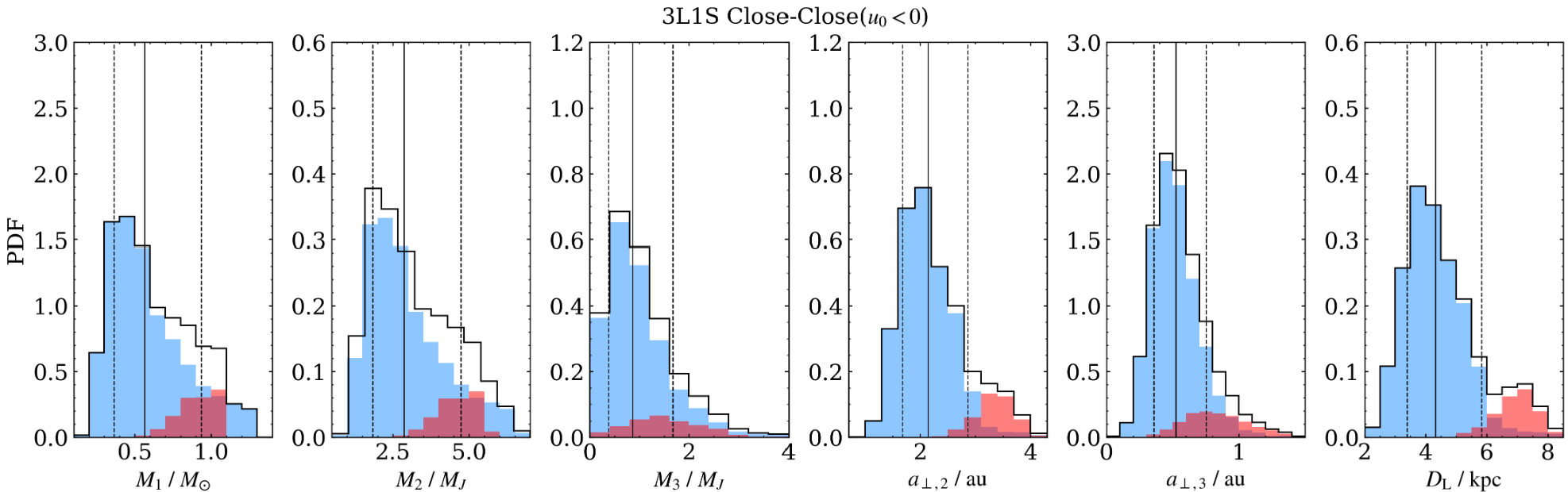}
    \includegraphics[width=0.8\textwidth]{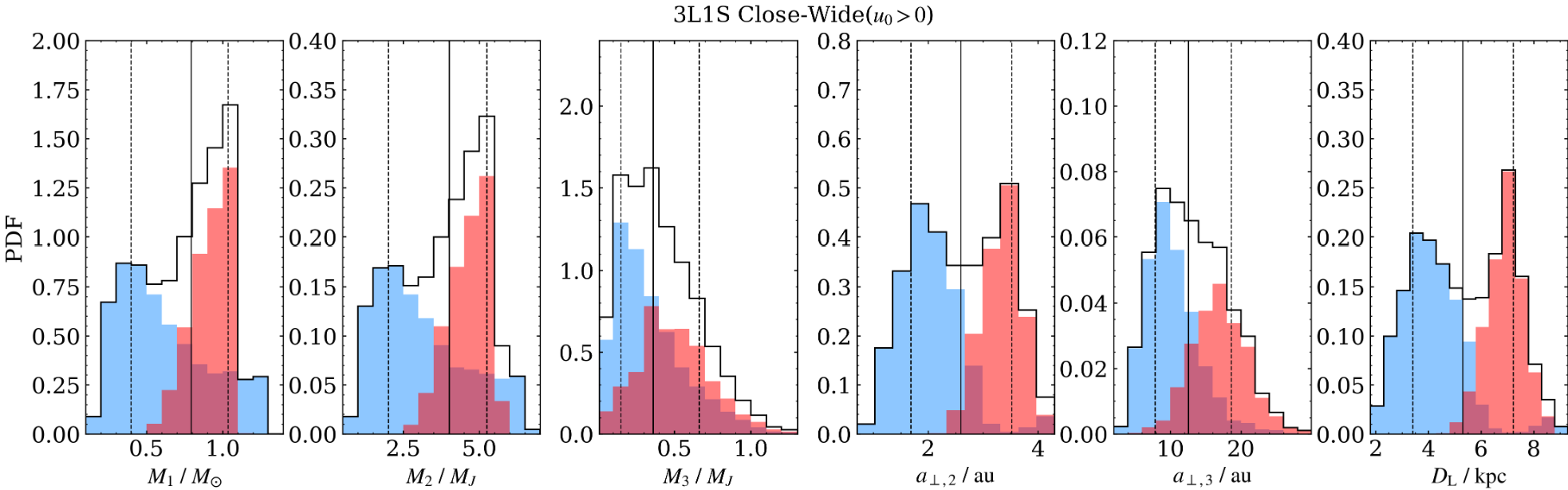}
    \includegraphics[width=0.8\textwidth]{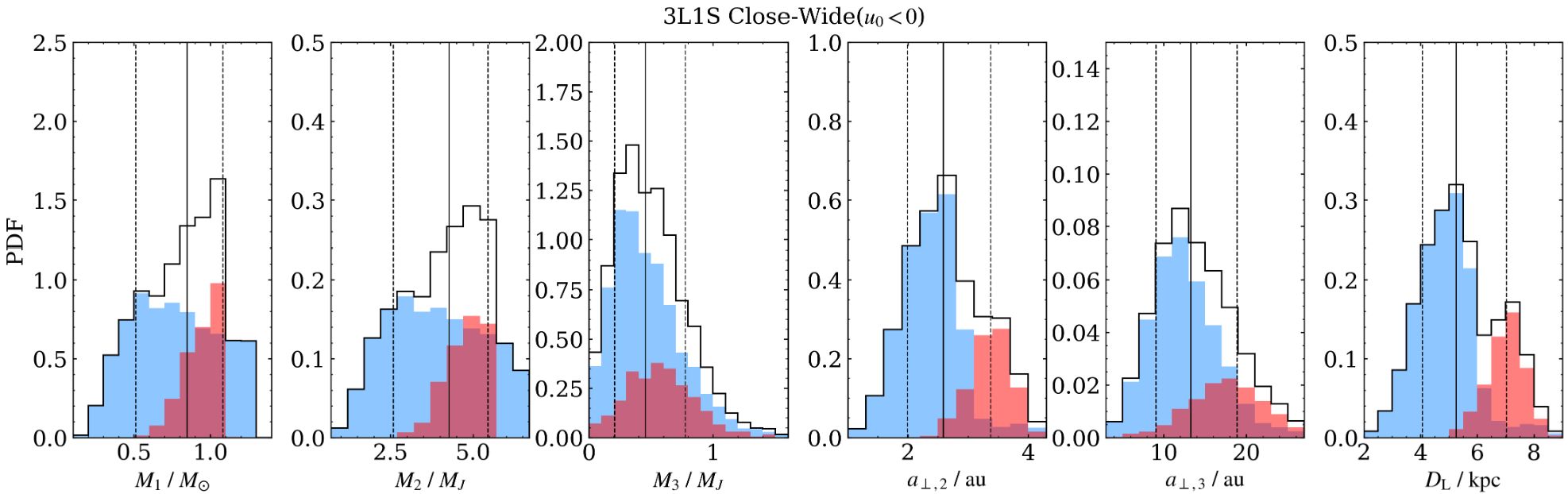}  
    \caption{Bayesian posterior distributions of the lens component masses ($M_1$ and $M_2$ for the 2L1S model, and $M_1$, $M_2$, and $M_3$ for the 3L1S model), the lens distance $D_{\rm L}$, and the projected planet-host separations ($a_{\perp 1}$ for the 2L1S model and $a_{\perp 1}$ and $a_{\perp 2}$ for the 3L1S model). In each panel, the black solid line indicates the median value, and the black dashed lines mark the 15.9\% and 84.1\% percentiles of the distribution. The corresponding numerical values are listed in Table~\ref{tab:phy}. The distributions for bulge and disk lenses are shown in red and blue, respectively, with the transparent histograms representing the combined distribution.}
\label{fig:phy}
\end{figure*}

In Table~\ref{tab:phy} and Figure \ref{fig:phy}, we show the posterior distributions of the lens physical parameters. Table~\ref{tab:phy} also presents the relative $\Delta\chi^2$ values between models, based on the probabilities from the Galactic model and the light-curve analysis. Among all models, the 3L1S ``Close-Wide'' ($u_0 > 0$) model has the highest probability and the other three models are disfavored by $\Delta\chi^2 \leq 2.6$. The second body is a super-Jupiter located at a projected separation of $\sim 2$ au from the host star. The third body is likely a Saturn-class planet, with a projected separation of either $\sim 0.4$ au and $\sim 13$ au for the ``Close-Close'' and ``Close-Wide'' models, respectively. The host star is most consistent with a K dwarf residing in the Galactic disk.

\section{Discussion}\label{sec:dis}

\subsection{A New Chanel for Two-planet Systems}\label{sec:31}

\begin{figure}
    \centering
    \includegraphics[width=0.47\textwidth]{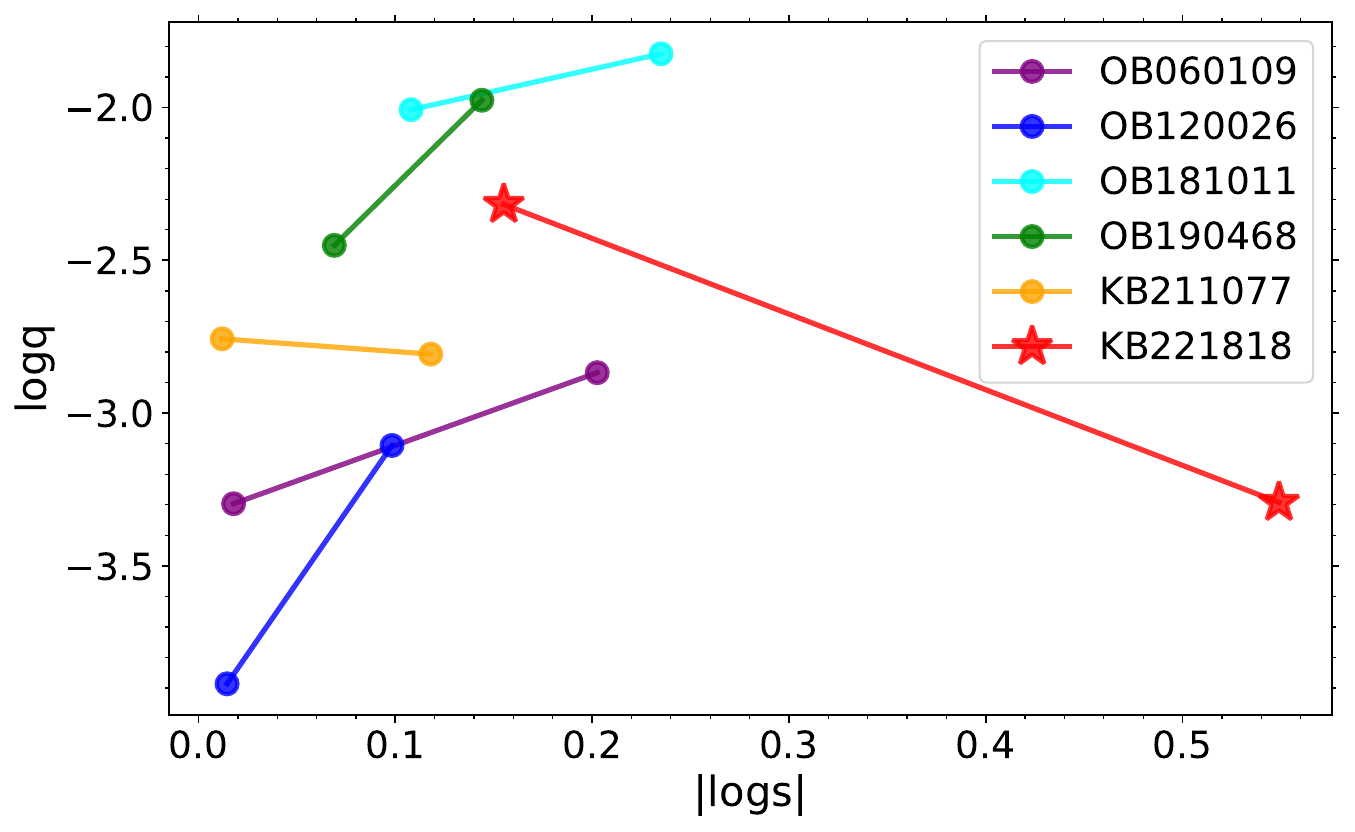}
    \caption{Distribution of $|\log s|$ versus $\log q$ for all six known two-planet microlensing systems. Different colors represent different systems, and the two planets in each system are connected by a line. The parameters shown are adopted from the best-fit 3L1S model for each system. Event names are abbreviations, e.g., \event\ to KB221818.}
\label{fig:logslogq}
\end{figure}

In this paper, we have presented the analysis of the sixth microlensing two-planet system, KMT-2022-BLG-1818L. Figure~\ref{fig:logslogq} shows the distribution of $|\log s|$ versus $\log q$ for all six known two-planet microlensing systems. Unlike the other five systems, in which both planets lie close to the Einstein ring (i.e., $|\log s| < 0.25$), the second planet in KMT-2022-BLG-1818L, designated KB221818Lc, lies much farther from the Einstein ring, with $|\log s| > 0.5$. Because the shear, which quantifies the perturbative strength of a companion relative to the primary lens \citep{CR1984}, is defined as $\gamma = q s^2$ and $\gamma = q / s^2$ for ``Close'' and ``Wide'' topologies, respectively, the signal of KB221818Lc is weaker than that of any other second planet. Our dense follow-up for HM events detected this subtle signal and broke the Close/Wide degeneracy for the second planet.

The large $|\log s|$ value of the second planet indicates that our follow-up program is capable of probing two-planet systems over a broader physical parameter space. In the five previously known microlensing two-planet systems, both planets are mostly located on Jupiter-like orbits. In contrast, for the present case, the ``Close'' third-body topology feature a Saturn-class planet at a projected separation of $\sim 0.4$ au from the host, corresponding to a Mercury-like orbit. Such configurations, with two gas giants on Mercury-like and Jupiter-like orbits, are more commonly found via other detection methods. A similar system is Kepler-539 \citep{Mancini2016}, in which the inner gas giant was detected via transits and the outer companion through transit timing variations. For the ``Wide'' third-body topology, the second planet is a Saturn-class planet at a projected separation of $\sim 13$ au. This makes it the lowest-mass planet at $>10$ au known in a multiple-planet system. The previous record was a $1.6~M_J$ planet at 12 au orbiting 47 UMa, discovered by radial velocity \citep{Gregory2010}. 

In addition to probing a broader region of physical parameter space, our analysis reveals a previously unidentified degeneracy within the 3L1S model, as well as a new degeneracy between the 3L1S and 2L1S models.

\subsection{A New Four-fold Degeneracy for 3L1S?}\label{sec:32}

\begin{figure*}
    \centering
    \includegraphics[width=0.47\textwidth]{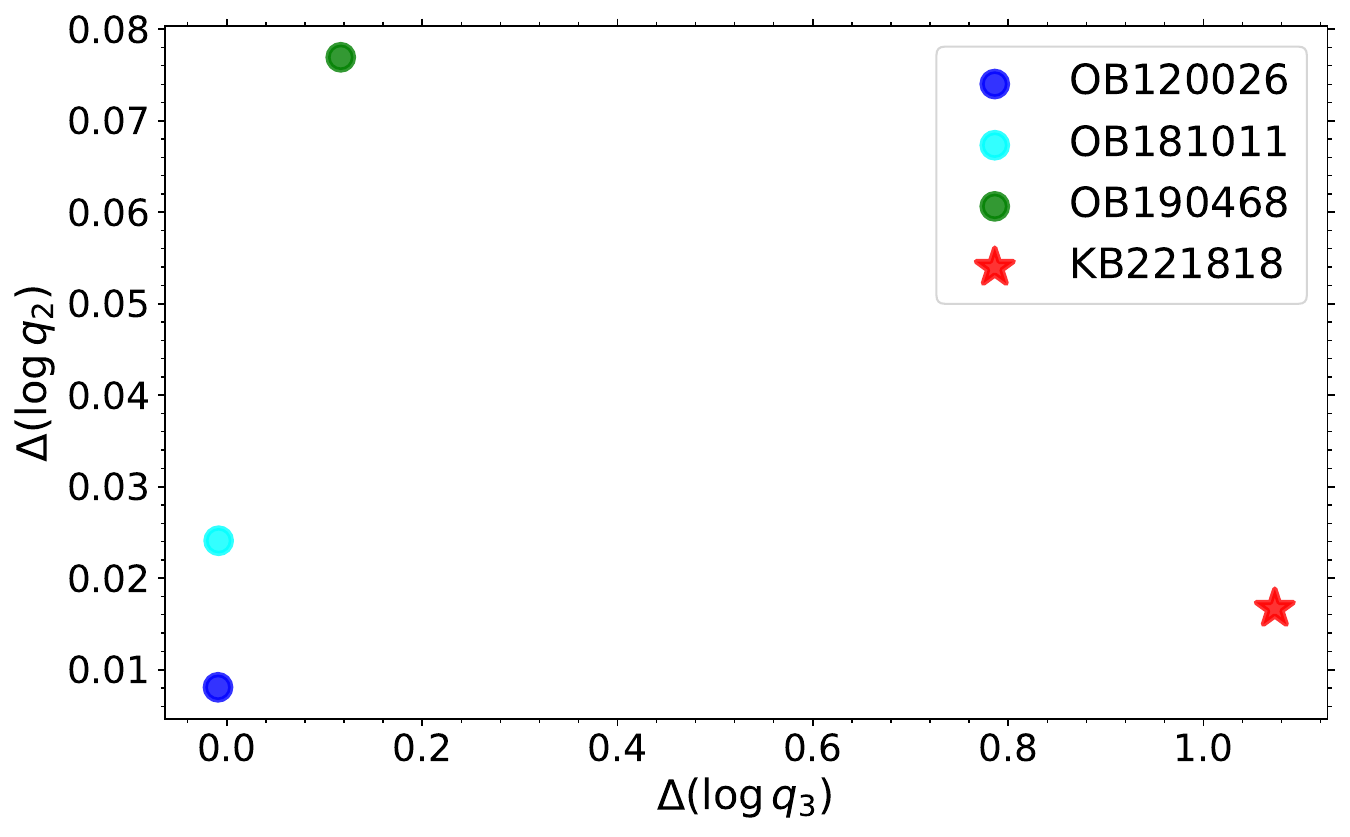}
    \includegraphics[width=0.47\textwidth]{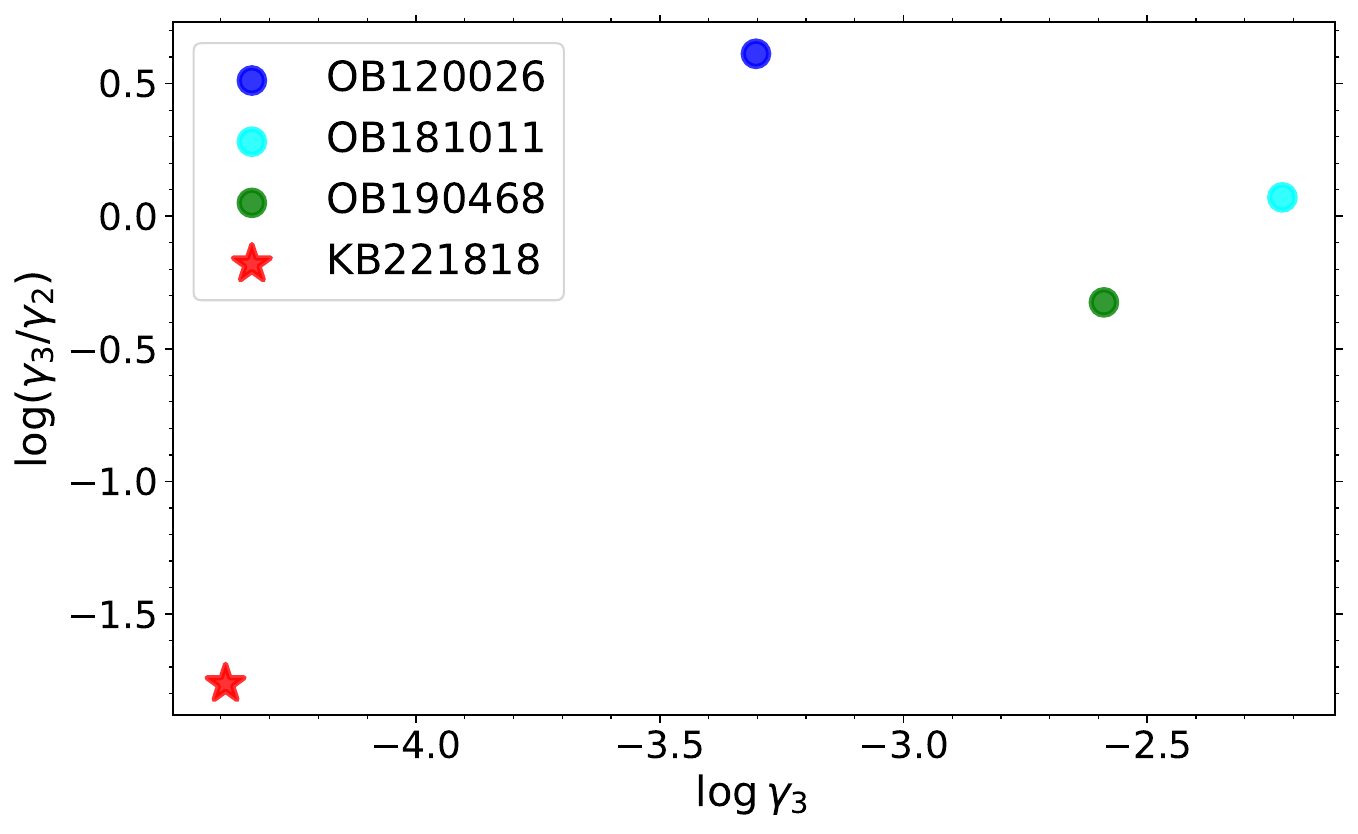}
    \caption{Both panels show distributions for the four known two-planet systems that exhibit a four-fold degeneracy. For the left panel, $\Delta(\log q_2)$ is defined as the mean $\log q_2$ for the ``Wide'' topology of the second body minus the mean $\log q_2$ for the ``Close'' topology. Similarly, $\Delta(\log q_3)$ is the mean $\log q_3$ for the ``Wide'' topology of the second body minus the mean $\log q_3$ for the ``Close'' topology. For the right panel, $\gamma_2$ and $\gamma_3$ are the shears of the second and third bodies, respectively, calculated using the parameters from the best-fit model of each event.}
\label{fig:dlogslogq}
\end{figure*}

The 3L1S model of \event\ exhibits a four-fold degeneracy, comprising the close/wide degeneracy for both the second and third bodies. This type of degeneracy is common in two-planet microlensing systems and has been identified in three previous two-planet events: OGLE-2012-BLG-0026L, OGLE-2018-BLG-1011L, and OGLE-2019-BLG-0468L. For these three systems, as well as KMT-2022-BLG-1818L, the left panel of Figure~\ref{fig:dlogslogq} displays the differences in mass ratio for the second ($\Delta(\log q_2)$) and third ($\Delta(\log q_3)$) bodies between the ``Close'' and ``Wide'' topologies of the second body. While the previously known systems show small variations in $\log q_3$ ($\Delta(\log q_3) < 0.12$), our event reveals a significantly larger difference of $\Delta(\log q_3) = 1.1$, marking the first instance for which the inferred properties of the third planet differ substantially between the degenerate topologies of the second planet. It is unclear whether this behavior is connected to the continuous degeneracy described by \cite{ref:song}.

The right panel of Figure~\ref{fig:dlogslogq} shows the shear induced by the second and third bodies, In the previously known systems with four-fold degeneracy, the shear differences between the second and third planets are modest, with $|\log \gamma_3/\gamma_2| < 0.6$. In contrast, our event displays a much larger shear difference of $\log \gamma_3/\gamma_2 = -1.7$. This may offer insight into the origin of the four-fold degeneracy shown in our system.

\subsection{A New Degeneracy between 3L1S and 2L1S}\label{sec:33}

In our analysis of \event, we uncover a previously unrecognized degeneracy between the 2L1S model with high-order effects and the 3L1S model. Specifically, the 2L1S model, incorporating the microlensing parallax and lens orbital motion effects, can reproduce subtle distortions in the light curve that are similarly produced by the 3L1S model through the presence of a third body. Although the 2L1S model is excluded for the current event based on $\Delta\chi^2 = 63.6$ and implausible physical parameters, this case raises two questions for future events. Can this degeneracy be broken, and can the presence of a third body be recognized?

For our case, if the follow-up observations during the anomaly had been less dense or less precise, the $\Delta\chi^2$ between the 2L1S and 3L1S models might not have been sufficient to confidently exclude the 2L1S model. Furthermore, if we hypothetically doubled the event timescale $\te$ while keeping all other parameters fixed, the ratio of transverse kinetic to potential energy, $\beta \propto \eta^2$ (see Equation~(\ref{equ:orbital})), would decrease to $\sim 0.2$, making the lens orbital configuration inferred from the 2L1S model appear physically reasonable. In such a scenario, one might incorrectly accept the 2L1S model as the true solution and never consider the 3L1S model. 

Prior to our event, this type of degeneracy had been identified in only three microlensing cases. In MACHO-97-BLG-41 \citep{Bennett1999, Albrow2000, Jung2013} and OGLE-2013-BLG-0723 \citep{OB130723_1, OB130723_2}, the light curves were initially interpreted as 3L1S events, but subsequent analyses revealed that 2L1S models with high-order effects provided better fits. In the case of KMT-2021-BLG-0322, the 3L1S and 2L1S models yielded comparably good fits, and the degeneracy could not be resolved. However, in all three cases, the 2L1S models involved binary lenses with stellar mass ratios. Our event represents the first instance in which this degeneracy has been identified in the 2L1S model with a planetary mass ratio.

From another perspective, physically implausible lens parameters inferred from a 2L1S model may serve as a clue pointing to the presence of a third body. This possibility was previously recognized in the analysis of KMT-2020-BLG-0414, and we refer the reader to Section 6.1 of \cite{KB200414}. 

With stable photometry and continuous coverage provided by upcoming space-based microlensing surveys, Roman \citep{MatthewWFIRSTI}, Earth 2.0 \citep{CMST,ET}, and CSST \citep{CSST_Wei}, the number of two-planet systems is expected to increase dramatically in the coming decade. A better understanding of the newly identified 3L1S degeneracy may aid in the search for 3L1S models. Moreover, our results suggest that future analyses should carefully examine the potential for degeneracies in which a seemingly adequate 2L1S fit may conceal a third body, or conversely, a 3L1S interpretation may instead arise from high-order effects in a 2L1S model. Recognizing and accounting for such degeneracies may be essential for accurately characterizing the architectures of microlensing two-planet systems.

\acknowledgments
H.L., J.Z., W.Zang, R.K., H.Y., S.M., Y.T., S.D., Z.L., Q.Q., and W.Zhu acknowledge support by the National Natural Science Foundation of China (Grant No. 12133005). W.Z. acknowledges the support from the Harvard-Smithsonian Center for Astrophysics through the CfA Fellowship. This research has made use of the KMTNet system operated by the Korea Astronomy and Space Science Institute (KASI) at three host sites of CTIO in Chile, SAAO in South Africa, and SSO in Australia. Data transfer from the host site to KASI was supported by the Korea Research Environment Open NETwork (KREONET). This research was supported by KASI under the R\&D program (Project No. 2025-1-830-05) supervised by the Ministry of Science and ICT. This research uses data obtained through the Telescope Access Program (TAP), which has been funded by the TAP member institutes. C.Han acknowledge the support from the Korea Astronomy and Space Science Institute under the R\&D program (Project No. 2025-1-830-05) supervised by the Ministry of Science and ICT.  H.Y. acknowledges support by the China Postdoctoral Science Foundation (No. 2024M762938).  This work makes use of observations from the Las Cumbres Observatory global telescope network. J.C.Y. and I.-G.S. acknowledge support from U.S. NSF Grant No. AST-2108414. Y.S. acknowledges support from BSF Grant No. 2020740. The MOA project is supported by JSPS KAKENHI Grant Number JP24253004, JP26247023,JP16H06287 and JP22H00153. Work by S.D. and Z.L. is supported by the China Manned Space Program with grant no. CMS-CSST-2025-A16. S.D. acknowledges the New Cornerstone Science Foundation through the XPLORER PRIZE. The authors acknowledge the Tsinghua Astrophysics High-Performance Computing platform at Tsinghua University for providing computational and data storage resources that have contributed to the research results reported within this paper.

\bibliography{Zang.bib}

\end{CJK*}
\end{document}